\documentstyle[aps,prb,epsf]{revtex}

\renewcommand{\epsilon}{\varepsilon}

\begin{document}
\draft 
\title{Classical critical behavior of spin models with long-range interactions}
\author{Erik Luijten\thanks{Electronic address: erik@tntnhb3.tn.tudelft.nl}
        and Henk W. J. Bl\"ote}
\address{Department of Physics, Delft University of Technology,
         Lorentzweg 1, 2628 CJ Delft, The Netherlands}
\date{\today}
\maketitle
\begin{abstract}
  We present the results of extensive Monte Carlo simulations of Ising models
  with algebraically decaying ferromagnetic interactions in the regime where
  classical critical behavior is expected for these systems.  We corroborate
  the values for the exponents predicted by renormalization theory for systems
  in one, two, and three dimensions and accurately observe the predicted
  logarithmic corrections at the upper critical dimension.  We give both
  theoretical and numerical evidence that above the upper critical dimension
  the decay of the critical spin--spin correlation function in finite systems
  consists of two different regimes.  For one-dimensional systems our estimates
  for the critical couplings are more than two orders of magnitude more
  accurate than existing estimates. In two and three dimensions we give, to our
  knowledge, the first results for the critical couplings.
\end{abstract}


\section{Introduction}
The critical behavior of Ising models with long-range interactions has
attracted much attention during the last three decades.  For the
one-dimensional case, some analytical results have been
obtained,\cite{ruelle68,dyson69a,dyson69b,thouless69,simon80,rogers81,%
  froehlich82,imbrie82,aizenman88,aizenman88b,imbrie88} as well as a number of
numerical results.  The numerical results apply to both inverse-square
interactions\cite{bhatta81,matvienko85,siu85,vigfusson86}
and general algebraically decaying interactions\@.\cite{rapaport68,dobson69,%
  nagle70,doman81,glumac89,monroe90,bergersen91,manieri92,monroe92,monroe94,%
  cannas95,pires96} Special mention deserves the work by Anderson, Yuval, and
Hamann,\cite{anderson69,yuval70,anderson70,anderson71} which greatly stimulated
the interest in spin chains with long-range interactions. They also developed a
renormalization-like approach to the one-dimensional inverse-square
model\@.\cite{anderson70,anderson71} Further renormalization-group studies of
this particular case are presented in
Refs.~\onlinecite{bhatta81,kosterlitz76,cardy81,bulgadaev84}.  A major
contribution was made by Fisher, Ma, and Nickel\cite{fisher72b} and
Sak,\cite{sak73} who obtained renormalization predictions for the critical
exponents of models of general dimensionality $d < 4$ with algebraically
decaying interactions (obtained independently by Suzuki {\em et
  al}.\cite{suzuki72}).  Other works concerning $d>1$ are two conjectures on,
respectively, the boundary between long-range and short-range behavior and the
boundary between classical (mean-field) and nonclassical behavior, both by
Stell,\cite{stell70} a (refuted) conjecture by Griffiths,\cite{griffiths70} a
rigorous confirmation of the upper critical dimension by Aizenman and
Fern\'andez,~\cite{aizenman88b} and a variational approach to the Ising model
with long-range interactions\@.\cite{wragg90} Furthermore, Monte Carlo
simulations have been carried out for one particular choice of the spin--spin
interaction in a two-dimensional model\@.~\cite{xu93} However, to our
knowledge, neither any further verifications of the renormalization predictions
nor any other results are available for higher-dimensional ($d>1$) models.  To
conclude this summary, we mention that the one-dimensional $q$-state Potts
model with long-range interactions has been studied
analytically,\cite{aizenman88,imbrie88} numerically,\cite{glumac93,cannas96}
and in a mean-field approximation on the Bethe lattice\@.\cite{bernardes94}

Why are these models interesting? In the first place from a fundamental point
of view: They enable us to study the influence of the interaction range on the
critical behavior. E.g., in one-dimensional systems long-range order is only
possible in the presence of spin--spin interactions which decay sufficiently
slowly. In the borderline (inverse-square) case, the 1D model displays a
remarkable behavior: At the critical temperature the order parameter exhibits a
finite jump (see Sec.~\ref{sec:rigorous}), but the free energy has an essential
singularity such that all thermal properties are smooth. In this sense, the
phase transition can be regarded as the one-dimensional analog of a
Kosterlitz--Thouless transition,\cite{kosterlitz73,kosterlitz74} although the
jump in the magnetization is not present there, as follows from the
Mermin--Wagner theorem\@.~\cite{merminwagner} Just as $d=2$ is the lower
critical dimension for the two-dimensional $XY$ model with short-range
interactions, $\sigma=1$ is a critical decay rate in a one-dimensional system
with interactions decaying as $r^{-(1+\sigma)}$, see
Ref.~\onlinecite{kosterlitz76}. With respect to higher-dimensional systems, we
note that the decay rate of van~der~Waals forces in realistic three-dimensional
systems is only slightly faster than at the boundary between short-range
(Ising-like) and long-range critical behavior. The question of criticality in
ionic systems, where the (screened) Coulomb interactions might lead to
effectively algebraically decaying interactions, appears still open to
debate\@.\cite{hafskjold,fisher94,folk95} It has also been claimed that
exponents in the long-range universality class have been observed
experimentally in a ferromagnetic phase transition\@.\cite{boxberg95} Recently,
it has been derived that critical fluctuations may give rise to long-range
Casimir forces (decaying much more slowly than van~der~Waals interactions)
between uncharged particles immersed in a critical fluid\@.\cite{burkhardt95}
Furthermore, it was shown by Anderson and Yuval\cite{anderson69,yuval70} that
the Kondo problem corresponds to a one-dimensional Ising model with a
combination of inverse-square and nearest-neighbor interactions. Yet another
application follows from Ref.~\onlinecite{bergersen91}, where it was shown that
random exchange (L\'evy-flight) processes can generate effective interactions
which decay algebraically. Hence, the universal critical properties of the
nonequilibrium steady state of these systems are those of the long-range
equilibrium Ising models studied in this paper. Finally, the realization that
the upper critical dimension can be varied by tuning the decay rate of the
interaction led to a special application of these models in
Ref.~\onlinecite{prl}.  Here, they were used to analyze a long-standing
controversy on the universality of the renormalized coupling constant above the
upper critical dimension.

In this article, we present accurate numerical results for Ising systems with
algebraically decaying interactions in one, two, and three dimensions.  Until
now, the long-range character of the spin--spin interactions has been the main
bottleneck for the examination of these systems by means of numerical methods
(and, in fact, also for their analytical solution).  All previously published
numerical results therefore rely on various extrapolations based on data for
small systems. However, the advent of a novel Monte Carlo algorithm\cite{ijmpc}
for the first time enabled us to efficiently simulate these systems. The high
accuracy of the results opens several perspectives: i) verification of the
renormalization predictions for the critical exponents; ii) accurate
observation of logarithmic corrections at the upper critical dimension; iii)
first estimates of the critical temperatures of two- and three-dimensional
systems with long-range interactions; iv) verification of previously obtained
estimates of the critical temperatures of one-dimensional systems, which in
addition implies a check on the various extrapolation methods that have been
developed; v) verification of predicted bounds on the critical temperatures;
vi) verification of a conjecture on the behavior of the critical temperature as
a function of the decay parameter.  Another problem one encounters in the
simulations is the large parameter space: The simulations for a set of
different temperatures and system sizes have to be repeated for a range of
values of the decay parameter and for $d=1,2,3$. The total computing time
dedicated to the results presented in this paper amounts to approximately two
CPU-years on a modern workstation.

The outline of this paper is as follows. In Sec.~\ref{sec:rigorous}, we sum
up the known rigorous results for the Ising chain with long-range interactions.
In Sec.~\ref{sec:fss}, we review the renormalization scenario of these
models and derive the finite-size scaling behavior of several quantities.
These include the corrections to scaling, both at and above the upper critical
dimension. Our numerical results are presented and analyzed in
Sec.~\ref{sec:numerical} and compared with previously obtained results.
Finally, we summarize our conclusions in Sec.~\ref{sec:conclusion}.  The
Appendix contains technical details concerning the application of the
long-range Monte Carlo algorithm to the models studied in this paper.

\section{Rigorous results for the one-dimensional case}
\label{sec:rigorous}
For the one-dimensional case, the Hamiltonian is given by
\begin{equation}
\label{eq:ham1d}
{\cal H} = \sum_{ij} J(i-j) s_i s_j \;,
\end{equation}
where the sum runs over all spin pairs.  We are particularly interested in
algebraically decaying interactions, i.e.\ $J(n) \propto n^{-\alpha}$. To
ensure that the energy of the system does not diverge, it is required that
$\alpha>1$.  In 1968, Ruelle\cite{ruelle68} rigorously proved the absence of
long-range order in a spin chain with ferromagnetic spin--spin couplings
$J(i-j)$ such that the sum
\begin{equation}
\label{eq:ruelle}
\sum_{n=1}^{N} n J(n)
\end{equation}
does not diverge in the limit $N \to \infty$. For algebraically decaying
interactions, this implies the absence of a phase transition for $\alpha>2$.
Shortly later, Dyson\cite{dyson69a} proved the {\em existence\/} of a phase
transition if the sums $\sum_{n=1}^{N} J(n)$ and $\sum_{n=1}^{N} (\log\log
n)\left[n^3 J(n) \right]^{-1}$ both converge, for positive and monotonically
decreasing $J(n)$. In particular, a phase transition occurs for $J(n) \propto
n^{-\alpha}$ with $1 < \alpha < 2$. This {\em partly\/} corroborated the
conjecture of Kac and Thompson,\cite{kac69} {\em viz.\/}\ that there is a phase
transition for $1 < \alpha \leq 2$. Furthermore, Dyson\cite{dyson69b} was (as
were---much later---also Rogers and Thompson\cite{rogers81}) able to replace
Ruelle's condition with a stronger one, which however still left the case
$\alpha=2$ undecided. This also holds for an even more stringent criterion by
Thouless,\cite{thouless69} who generalized the argument of Landau and
Lifshitz\cite{landau} for the absence of a phase transition in an Ising chain
with short-range interactions. However, Thouless argued on entropic grounds
that {\em if\/} a phase transition exists for $\alpha=2$, the magnetization
must have a {\em discontinuity\/} at the transition point. This was later
dubbed the ``Thouless effect'' by Dyson, who proved it to occur in the closely
related hierarchical model\@.\cite{dyson71} Simon and Sokal made Thouless'
argument partially rigorous,\cite{simon80} but later Aizenman {\em et
  al.}\cite{aizenman88} showed that, although a discontinuity in the order
parameter {\em is indeed present\/} if there is a phase transition, his
argument does {\em not\/} account for this. Namely, Thouless had assumed that
the spin--spin correlation function $\langle s_0 s_r \rangle - \langle s_0
\rangle \langle s_r \rangle$ vanishes in the limit $r \to \infty$, whereas
actually the critical exponent $\eta$ is equal to 1 in this case.  Meanwhile,
Fr\"ohlich and Spencer\cite{froehlich82} had been able to rigorously prove the
{\em existence\/} of a phase transition in the borderline case and thus to
corroborate the Kac--Thompson conjecture for $\alpha=2$ as well.  Another
interesting point is the rigorous proof for the existence of an intermediate
ordered phase in the one-dimensional model with inverse-square interactions,
where the two-point correlation function exhibits power-law decay with an
exponent which varies continuously in a finite temperature range below the
critical temperature\@.\cite{imbrie88}

\section{Finite-size analysis of the critical behavior}
\label{sec:fss}
Already in a very early stage of the history of the $\epsilon$-expansion,
Fisher, Ma, and Nickel analyzed the critical behavior of $d$-dimensional
systems ($d<4$) with long-range interactions decaying as $r^{-(d+\sigma)}$,
with $\sigma>0$\@.\cite{fisher72b} They concluded that the upper critical
dimension is given by $d_{\rm u}=2\sigma$, as was previously conjectured by
Stell\cite{stell70} and later rigorously proven by Aizenman and
Fern\'andez\@.\cite{aizenman88b} For more slowly decaying interactions, $0 <
\sigma < d/2$, the critical behavior is classical, whereas the critical
exponents assume nonclassical, continuously varying values for $d/2 < \sigma <
2$.  For $\sigma>2$ they take their short-range (Ising) values.
Sak,\cite{sak73} however, found that already for $\sigma>2-\eta_{\rm sr}$ the
critical behavior is Isinglike, where $\eta_{\rm sr}$ denotes the exponent
$\eta$ in the corresponding model with short-range interactions.  In this
article we concentrate on the classical range, for which we have performed
extensive Monte Carlo simulations of spin models in $d=1,2,3$. The nonclassical
range will be the subject of a future article\@.\cite{nonclass}

We briefly outline the renormalization scenario for these models, in order to
derive the finite-size scaling relations required to analyze the numerical
data. We start from the following Landau--Ginzburg--Wilson Hamiltonian in
momentum space,
\begin{equation}
{\cal H}(\phi_{{\bf k}})/k_{\rm B}T =
  \frac{1}{2}\sum_{{\bf k}} \left( j_\sigma k^\sigma + j_2 k^2 + r_0 \right)
  \phi_{{\bf k}}\phi_{-{\bf k}} +
  \frac{u}{4N} \sum_{{\bf k}_1}\sum_{{\bf k}_2}\sum_{{\bf k}_3}
  \phi_{{\bf k}_1}\phi_{{\bf k}_2}\phi_{{\bf k}_3}
  \phi_{-{\bf k}_1-{\bf k}_2-{\bf k}_3} -
  h\sqrt{\frac{N}{2}}\phi_{{\bf k}={\bf 0}} \;.
\label{eq:lgw}
\end{equation}
The $j_\sigma k^\sigma$ term arises from the Fourier transform of the
interactions decaying as $r^{-(d+\sigma)}$. The $j_2 k^2$ term normally
representing the short-range interactions is included because it will appear
anyway in the renormalization process and will compete with the long-range
term\@.\cite{sak73} Under a renormalization transformation with a rescaling
factor $b=e^l$, the term $j_\sigma k^\sigma$ is transformed into $j_\sigma
k'^\sigma$, with ${\bf k}' = {\bf k}b$. To keep the coefficient of the
$k^\sigma$ term fixed, we rescale the field $\phi_{{\bf k}}$ to
$\phi'_{{\bf k}'}=b^{-\sigma/2}\phi_{{\bf k}}$.  Thus, the coefficient of the
$k^2$ term decreases as $b^{\sigma-2}$ and the coefficient of the $\phi^4$ term
changes proportional to $b^{2\sigma-d}$. Hence, the Gaussian fixed point
dominates the renormalization flow for $\sigma < d/2$, which is the situation
studied in this paper.

For the sake of generality we treat here the case of an $n$-component order
parameter with ${\rm O}(n)$ symmetry.  The renormalization equations are then
given by
\begin{mathletters}
\begin{eqnarray}
\label{eq:rg-r0}
\frac{d r_0}{dl} &=& \sigma r_0 + a(n+2) u (c - r_0) \;,\\
\label{eq:rg-u}
\frac{d u}{dl}   &=& \epsilon u - a(n+8) u^2         \;,
\end{eqnarray}
\end{mathletters}%
where $(n+2)$ and $(n+8)$ are the usual factors arising from the tensorial
structure of the interaction part of the Hamiltonian and $\epsilon=2\sigma-d$.
These equations are not complete to second order, because the ${\cal O}(u^2)$
term is missing in Eq.~(\ref{eq:rg-r0}).

We first consider the case $\epsilon<0$.  The solution of the second equation
is given by
\begin{equation}
\label{eq:sol-u}
u(l) = \bar{u} e^{\epsilon l}
       \frac{1}{1+\bar{u}\frac{a(n+8)}{\epsilon} (e^{\epsilon l}-1)} \;,
\end{equation}
where $\bar{u}$ denotes the value of $u$ at $l=0$. This yields, to leading
order in $u$, the following solution for the first equation,
\begin{equation}
\label{eq:sol-r0}
r_0(l) = \left[\bar{r}_0 + ac(n+2)\bar{u}/(d-\sigma)\right] e^{\sigma l}
         \left[ \frac{1}{1+\frac{a(n+8)}{\epsilon}\bar{u}(e^{\epsilon l}-1)}
         \right]^{(n+2)/(n+8)}
         -\frac{ac(n+2)\bar{u} e^{\epsilon l} / (d-\sigma)}{1 +
          \frac{a(n+8)}{\epsilon}\bar{u}(e^{\epsilon l}-1)}\;,
\end{equation}
with $\bar{r}_0 \equiv r_0(l=0)$.  The first factor between square brackets is
proportional to the reduced temperature $t \equiv (T-T_{\rm c})/T_{\rm c}$ and
the last term is the so-called shift of the critical temperature.  The factors
$[1+a(n+8)\bar{u}(e^{\epsilon l}-1)/\epsilon]^{-1}$ in Eqs.~(\ref{eq:sol-u})
and~(\ref{eq:sol-r0}) are higher-order corrections in $u$.  Under successive
renormalization transformations, $u$ approaches the value $u^*=0$ and the
Gaussian fixed point $(0,0)$ is thus indeed stable. The pertinent
renormalization exponents are: $y_{\rm t}=\sigma$, $y_{\rm h}=(d+\sigma)/2$,
and $y_{\rm i}=2\sigma-d$.

At $\epsilon=0$, the Gaussian fixed point becomes marginally stable. Solving
Eq.~(\ref{eq:rg-u}) leads to
\begin{equation}
u^{\rm uc}(l) = \frac{\bar{u}}{1 + a(n+8)\bar{u}l} \;,
\end{equation}
where the superscript ``${\rm uc}$'' indicates that we are operating at the
upper critical dimension.  This solution can be used to solve, again to leading
order in $u$, Eq.~(\ref{eq:rg-r0}), yielding
\begin{equation}
\label{eq:log-r0}
r_0^{\rm uc}(l) = \left[\bar{r}_0 + ac(n+2)\bar{u}/(d/2)\right] e^{\sigma l}
         \left[\frac{1}{1+a(n+8)\bar{u}l}\right]^{(n+2)/(n+8)}
         - \frac{ac(n+2)\bar{u}/(d/2)}{1 + a(n+8)\bar{u}l}
\end{equation}
or, in terms of the rescaling factor $b$,
\begin{equation}
\label{eq:log-r0-b}
r_0^{\rm uc} = \left[\bar{r}_0 + ac(n+2)\bar{u}/(d/2)\right] b^{\sigma}
      \left[\frac{1}{1+a(n+8)\bar{u}\ln b}\right]^{(n+2)/(n+8)}
      - \frac{ac(n+2)\bar{u}/(d/2)}{1+a(n+8)\bar{u}\ln b} 
      \;.
\end{equation}
Since $\sigma$ is fixed at $d/2$ the factor $d/2$ in the last term
is identical to the corresponding factor $(d-\sigma)$ in
Eq.~(\ref{eq:sol-r0}).  Further comparison of Eqs.~(\ref{eq:sol-r0})
and~(\ref{eq:log-r0}) shows that above the upper critical dimension the leading
shift of the critical temperature is proportional to $b^{\epsilon}$, whereas
this factor vanishes at the upper critical dimension itself and the factor
$(e^{\epsilon l}-1)/\epsilon$ in the second-order correction turns into a $\ln
b$ term, yielding a logarithmic shift of the form $1/(A \ln b + B)$.

{}From the solutions of the renormalization equations we can derive the scaling
behavior of the free energy and of (combinations of) its derivatives.  For the
case $\epsilon<0$ the free energy density $f$ scales, to leading order, as
\begin{equation}
\label{eq:f-scale}
f(t,h,u,1/L) = b^{-d}f\left( b^{y_{\rm t}} \left[ t+\tilde{\alpha}ub^{y_{\rm
    i}-y_{\rm t}} \right ], b^{y_{\rm h}}h, b^{y_{\rm i}}u, b/L \right) + g \;,
\end{equation}
where $\tilde{\alpha}=-ac(n+2)/(d-\sigma)$ and we have included a finite-size
field $L^{-1}$. $g$ denotes the analytic part of the transformation. We
abbreviate the first term on the right-hand side as $b^{-d}f(t',h',u',b/L)$.
However, we must take into account the fact that, for $T \leq T_{\rm c}$, the
free energy is singular at $u=0$. This makes $u$ a so-called {\em dangerous\/}
irrelevant variable; see, e.g., Ref.~\onlinecite{privfish}. As discussed in
Ref.~\onlinecite{prl}, the correct finite-size scaling properties are obtained
by setting $b=L$ and making the substitution $\phi'=\phi/u'^{1/4}$. This leads
to a new universal function, $\tilde{f}$, with
\begin{equation}
\label{eq:f-subst}
f(t',h',u',1) + \bar{g} = \tilde{f}(\tilde{t},\tilde{h})\;,
\end{equation}
where $\tilde{t}=t'/u'^{1/2}$ and $\tilde{h}=h'/u'^{1/4}$. The analytic part of
the transformation also contributes to the singular dependence of the free
energy on $t$ (see, e.g., Ref.~\onlinecite[Ch.~VI, \S~3]{ma-book}): Despite
the regularity of this term in each single renormalization step, the infinite
number of steps still leads to the build-up of a singularity.  This
contribution, denoted by $\bar{g}$, is absorbed in $\tilde{f}$ as well. Setting
$b=L$ and combining Eqs.~(\ref{eq:f-scale}) and~(\ref{eq:f-subst}) yields
\begin{mathletters}
\begin{eqnarray}
f\left(t,h,u,\frac{1}{L}\right)
   &=& L^{-d}\tilde{f}\left(L^{y_{\rm t}-y_{\rm i}/2} \frac{1}{u^{1/2}}
       \left[t + \tilde{\alpha} uL^{y_{\rm i}-y_{\rm t}}\right],
       L^{y_{\rm h}-y_{\rm i}/4}\frac{h}{u^{1/4}} \right) \\
   &=& L^{-d}\tilde{f}\left(L^{y_{\rm t}^*} \frac{1}{u^{1/2}}
       \left[t + \tilde{\alpha} uL^{y_{\rm i}-y_{\rm t}}\right],
       L^{y_{\rm h}^*} \frac{h}{u^{1/4}} \right) \;.
\label{eq:f-fss}
\end{eqnarray}
\end{mathletters}%
Here, we have introduced the exponents $y_{\rm t}^* \equiv y_{\rm t}-y_{\rm
  i}/2=d/2$ and $y_{\rm h}^* \equiv y_{\rm h}-y_{\rm i}/4=3d/4$. The
corresponding critical exponents indeed assume their fixed, classical values;
$\alpha=0$, $\beta=1/2$, $\gamma=1$, $\delta=3$. The exponent $\gamma$ is
singled out here as a special case; even without taking into account the
modification of $y_{\rm t}$ and $y_{\rm h}$ due to the dangerous irrelevant
variable one obtains the classical value $\gamma=1$.  Since the correlation
length exponent $\nu=1/y_{\rm t}$ (it is not affected by the singular
dependence of the free energy on $u$), we see that hyperscaling is violated,
which is a well-known result for systems above their upper critical
dimension\@.\cite{privfish} The rescaling of the pair-correlation function $g$
(decaying proportional to $1/r^{d-2+\eta}$) relates the exponent $\eta$ to the
rescaling factor of the field, yielding $\eta=2-\sigma$. Note that this
contrasts with the short-range case ($\sigma=2$), where $\eta$ assumes its
mean-field value for all dimensionalities $d \geq 4$.  This implies that direct
experimental measurement of either $\nu$ or $\eta$ offers a way to discern
whether the interactions in a system are mean-field-like ($\nu=1/2$, $\eta=0$)
or have the form of a slowly decaying power-law. {\em Below\/} the upper
critical dimension, however, the finite-size scaling behavior of the spin--spin
correlation function is (apart from a volume factor) identical to that of the
magnetic susceptibility $\chi$.  This relation yields a contradiction above the
upper critical dimension, since $\chi$ depends on the scaled combination
$tL^{y_{\rm t}^*}$, instead of $tL^{y_{\rm t}}$. Indeed, the susceptibility
diverges as $t^{-\gamma}$ and the finite-size behavior of $\chi$ is thus
$\chi_L \propto L^{\gamma y_{\rm t}^*} = L^{d/2}$, corresponding to $g \propto
L^{-d/2}$.  On the other hand, if one assumes that the finite-size behavior of
the correlation function is identical to its large-distance behavior, one
expects that $g \propto L^{-(d-2+\eta)} = L^{-(d-\sigma)}$.  Only at the upper
critical dimension, $d_{\rm u}=2\sigma$, these two predictions coincide. We
will return to this point at the end of this section.  Furthermore, we will
examine the behavior of the spin--spin correlation function in
Sec.~\ref{sec:numerical}.

At the upper critical dimension itself, i.e.\ at $\epsilon=0$, the free energy
density scales as
\begin{mathletters}
\begin{eqnarray}
\label{eq:f-scale-du}
\lefteqn{f\left(t,h,u,\frac{1}{L}\right)} \nonumber\\
  &=& b^{-d}f\left( 
      \frac{b^{y_{\rm t}}}{(1+\tilde{\beta}u\ln b)^{(n+2)/(n+8)}} 
      \left[ t+\tilde{\alpha}b^{-y_{\rm t}}
      \frac{u}{(1+\tilde{\beta}u \ln b)^{6/(n+8)}} \right ], b^{y_{\rm h}}h,
      \frac{u}{1+\tilde{\beta}u \ln b}, \frac{b}{L} \right) + g \\
  &=& L^{-d}\tilde{f}\left( 
      \frac{L^{y_{\rm t}}}{(1+\tilde{\beta}u\ln L)^{(n+2)/(n+8)-1/2}} 
      \frac{1}{u^{1/2}}
      \left[ t+\tilde{\alpha}L^{-y_{\rm t}}
             \frac{u}{(1+\tilde{\beta}u \ln L)^{6/(n+8)}}
      \right ],
      L^{y_{\rm h}} \frac{h}{u^{1/4}} [1+\tilde{\beta}u \ln L]^{1/4} 
      \right) \;,
\label{eq:f-fss-du}
\end{eqnarray}
\end{mathletters}%
where $\tilde{\beta}=a(n+8)$ and we have set $b=L$ in the last line.  $u$ is
now a marginal variable and although we again have to perform the substitution
$\phi \to \phi'$ (the Gaussian fixed point is marginally stable), the exponents
$y_{\rm t}$ and $y_{\rm h}$ coincide with $y_{\rm t}^*$ and $y_{\rm h}^*$,
respectively, because $y_{\rm i}$ vanishes. Thus, the scaling
relations~(\ref{eq:f-fss}) and~(\ref{eq:f-fss-du}) differ to leading order only
in the logarithmic factors arising in the arguments of $\tilde{f}$.

As usual, the finite-size scaling relations are now found by taking derivatives
of the free energy density with respect to the appropriate scaling fields.  In
the Monte Carlo simulations we have sampled the second and the fourth moment of
the magnetization density, the dimensionless amplitude ratio $Q \equiv \langle
m^2 \rangle^2 / \langle m^4 \rangle$ (which is directly related to the Binder
cumulant~\cite{bindercum}), and the spin--spin correlation function over half
the system size (for even system sizes). The second moment of the magnetization
density is (apart from a volume factor) equal to the second derivative of the
free energy density with respect to $h$,
\begin{equation}
\langle m^2 \rangle = L^{-d} \frac{\partial^2 f}{\partial h^2}(t,h,u,1/L)
   = L^{2y_{\rm h}^*-2d} u^{-1/2} \tilde{f}^{(2)}\left( L^{y_{\rm t}^*}
     \frac{\hat{t}}{u^{1/2}}, L^{y_{\rm h}^*}\frac{h}{u^{1/4}} \right) \;,
\label{eq:m-fss}
\end{equation}
where $\tilde{f}^{(2)}$ stands for the second derivative of $\tilde{f}$ with
respect to its second argument and $\hat{t} \equiv t + \tilde{\alpha} u
L^{y_{\rm i}-y_{\rm t}}$.  At $\epsilon = 0$, logarithmic factors do arise not
only in the arguments of $\tilde{f}^{(2)}$, but also in the prefactor,
\begin{eqnarray}
\langle m^2 \rangle &=& L^{2y_{\rm h}-2d} 
   \left( \frac{1+\tilde{\beta}u \ln L}{u} \right)^{1/2} \nonumber\\
   & & \times \tilde{f}^{(2)}\left( 
      \frac{L^{y_{\rm t}}}{(1+\tilde{\beta}u\ln L)^{(n+2)/(n+8)-1/2}}
      \frac{1}{u^{1/2}} 
      \left[t+\tilde{\alpha}L^{-y_{\rm t}} 
            \frac{u}{(1+\tilde{\beta}u \ln L)^{6/(n+8)}}
      \right ],
      L^{y_{\rm h}} \frac{h}{u^{1/4}} [1+\tilde{\beta}u \ln L]^{1/4}
      \right) \;.
\label{eq:m-fss-du}
\end{eqnarray}
For the fourth magnetization moment similar expressions hold and in the
amplitude ratio~$Q$ all prefactors divide out, both for $\epsilon < 0$ and
$\epsilon = 0$. Thus we find that the ratio~$Q$ is given by a universal
function~$\tilde{Q}$,
\begin{equation}
\label{eq:q-scalfunc}
Q_L(T) = \tilde{Q}\left( L^{y_{\rm t}^*} \frac{\hat{t}}{u^{1/2}} \right)
         + q_1 L^{d-2y_{\rm h}^*} + \cdots \;,
\end{equation}
where we have omitted the $h$ dependence of~$\tilde{Q}$, since we are only
interested in the case $h=0$. The additional term proportional to $q_1$ arises
from the $h$ dependence of the analytic part of the free energy~\cite{ic3d} and
the ellipsis stands for higher powers of $L^{d-2y_{\rm h}^*}$ (faster-decaying
terms).  At $\epsilon = 0$, $\hat{t}$ must be replaced by the first argument
within square brackets in Eq.~(\ref{eq:f-fss-du}), multiplied by the factor
$(1+\tilde{\beta}u\ln L)^{1/2-(n+2)/(n+8)}$.  Finally, we may derive the
finite-size scaling behavior of the spin--spin correlation function $g({\bf
  r})$ by differentiating the free energy density to two {\em local\/} magnetic
fields, which couple to the spins at positions ${\bf 0}$ and ${\bf r}$,
respectively, and assuming that the finite-size behavior is identical to the
$r$ dependence of $g$.  If we do not take into account the dangerous irrelevant
variable mechanism, we find $g \propto L^{2y_{\rm h}-2d} = L^{-(d-\sigma)}$,
just as we found before from $\eta=2-\sigma$. However, replacing $y_{\rm h}$ by
$y_{\rm h}^*$ yields $g \propto L^{-d/2}$, in agreement with the $L$ dependence
of the magnetic susceptibility. This clarifies the difference between the two
predictions: At short distances (large wave vectors), the $j_\sigma k^\sigma
\phi_{\bf k}\phi_{-{\bf k}}$ term will be the dominant term in the
Landau--Ginzburg--Wilson Hamiltonian and there is no ``dangerous'' dependence
on $u$. Hence, the finite-size behavior of the spin--spin correlation function
will be given by $L^{-(d-2+\eta)}$. For ${\bf k}={\bf 0}$, the coefficient of
the $\phi^2$ term vanishes and thus the $u\phi^4$ term is required to act as a
bound on the magnetization. To account for this singular dependence on $u$, we
rescale the field, which implies that $y_{\rm h}$ is replaced by $y_{\rm h}^*$
and $g$ scales as $L^{2y_{\rm h}^*-2d}$. In a finite system, the wave vectors
assume discrete values, ${\bf k}=(n_x,n_y,n_z)2\pi/L$, and thus it is easily
seen that even for the lowest nonzero wave vectors $j_\sigma k^\sigma
\phi_{{\bf k}}\phi_{-{\bf k}}$ constitutes the dominant bounding term on the
magnetization. Namely, the coefficient of the $\phi^4$ term contains a volume
factor $L^{-d}$ [cf.\ Eq.~(\ref{eq:lgw})] and this term is thus (above the
upper critical dimension) a higher-order contribution decaying as
$L^{2\sigma-d}$.

\section{Numerical results and comparison with earlier results}
\label{sec:numerical}
\subsection{Simulations}
We have carried out Monte Carlo simulations for systems described by the
Hamiltonian
\begin{equation}
 {\cal H}/k_{\rm B}T = 
            - \sum_{\langle i j \rangle} J(|{\bf r}_i-{\bf r}_j|) s_i s_j \;,
\end{equation}
where the sum runs over all spin pairs and periodic boundaries were employed.
The precise form of the (long-range) spin--spin interaction $J(r)$ as used in
the simulations was chosen dependent on the dimensionality. For $d=1$ we have
followed the conventional choice $J(r)=K/r^{d+\sigma}$ (with {\em discrete\/}
values for $r$), as this allows us to compare {\em all\/} our results
(including nonuniversal quantities) to previous estimates. However, as
explained in Ref.~\onlinecite{ijmpc} and the Appendix, this discrete form
requires the construction of a look-up table, which becomes inefficient for
higher dimensionalities. For $d=2$ we have thus applied an interaction which is
the integral of a continuously decaying function,
\begin{equation}
J(|{\bf r}|) = K \int_{r_x-\frac{1}{2}}^{r_x+\frac{1}{2}} {\rm d}x
         \int_{r_y-\frac{1}{2}}^{r_y+\frac{1}{2}} {\rm d}y \;
         r^{-(d+\sigma)} \;,
\label{eq:2d-j}
\end{equation}
where ${\bf r}=(r_x,r_y)$ and $r=|{\bf r}|$. In $d=3$ the corresponding volume
integral was used for $J(r)$. This modification of the interaction does only
change nonuniversal quantities like the critical temperature, but should not
influence the universal critical properties like the critical exponents and
dimensionless amplitude ratios, since the difference between the continuous and
the discrete interaction consists of faster decaying terms that are irrelevant
according to renormalization theory. Details concerning the simulations can be
found in the Appendix.

The following system sizes have been examined: chains of length $10 \leq L \leq
150000$, square systems of linear size $4 \leq L \leq 240$, and cubic systems
of linear size $4 \leq L \leq 64$.  At the upper critical dimension simulations
for even larger systems have been carried out in order to obtain accurate
results from the analyses: $L=300000$ in $d=1$ and $L=400$ in $d=2$.  (I.e., in
terms of numbers of particles the largest system size for $d=2$ is considerably
smaller than for $d=1$ and $d=3$.)  For the simulations we used a new cluster
algorithm for long-range interactions\@.\cite{ijmpc} This algorithm is ${\cal
  O}(L^{d+z})$ times faster than a conventional Metropolis algorithm, where $z$
is the dynamical critical exponent. For systems displaying mean-field-like
critical behavior, we expect $z=d/2$ and the efficiency gain in our simulations
is thus of the order of $10^8$ for the largest system sizes.  For each data
point we have generated between $10^6$ and $4 \times 10^6$ Wolff clusters.


\subsection{Determination of the critical temperatures, the amplitude
  ratio~$Q$, and the thermal exponent}
The critical couplings $K_{\rm c}$ of these systems have been determined using
an analysis of the amplitude ratio~$Q$. The finite-size scaling analysis was
based on the Taylor expansion of Eq.~(\ref{eq:q-scalfunc}), which for
$\epsilon<0$ reads:
\begin{equation}
\label{eq:q-taylor}
 Q_L(T) = Q + p_1 \hat{t}L^{y_{\rm t}^*} + p_2 \hat{t}^2 L^{2y_{\rm t}^*}
          + p_3 \hat{t}^3 L^{3y_{\rm t}} + \cdots
          + q_1 L^{d-2y_{\rm h}^*} + \cdots
          + q_3 L^{y_{\rm i}} + \cdots \;.
\end{equation}
The term proportional to $\tilde{\alpha}$ in $\hat{t}$ yields a contribution
$q_2 L^{y_{\rm i}/2} = q_2 L^{\sigma-d/2}$ and the term $q_3 L^{y_{\rm i}}$
comes from the denominator in Eq.~(\ref{eq:sol-u}). The coefficients $p_i$ and
$q_i$ are nonuniversal. In addition to the corrections to scaling in
Eq.~(\ref{eq:q-taylor}) we have also included higher powers of $q_3 L^{y_{\rm
    i}}$, which become particularly important when $y_{\rm i}$ is small (i.e.\ 
when $\sigma$ is close to $d/2$), higher powers of $q_1 L^{d-2y_{\rm h}^*} =
q_1 L^{-d/2}$, and the crossterm proportional to $L^{y_{\rm t}^* + y_{\rm i}}$.

All analyses were carried out on the same data set as used in
Ref.~\onlinecite{prl}, to which several data points have been added for most
values of~$\sigma$.  First, we have only kept fixed the exponents in the
correction terms, $y_{\rm i}$ and $y_{\rm h}^*$.  The corresponding estimates
for $Q$ and~$y_{\rm t}^*$ are shown in the third and fourth column of
Table~\ref{tab:q_yt_fit}. One observes that the Monte Carlo results for both
$Q$ and~$y_{\rm t}^*$ are in quite good agreement with the renormalization
predictions~\cite{bzj,prl} $Q=8\pi^2/\Gamma^4(\frac{1}{4}) = 0.456947$\ldots\ 
and $y_{\rm t}^*=d/2$. However, the uncertainties in the estimates increase
considerably with increasing $\sigma$, because the leading irrelevant exponent
becomes very small. An exception is the relatively large uncertainty in $y_{\rm
  t}^*(d=1,\sigma=0.2)$, which originates from the fact that the Monte Carlo
data were taken in a rather narrow temperature region around the critical
point.  Furthermore, an accurate simultaneous determination of $Q$ and~$y_{\rm
  t}^*$ is very difficult, because of the correlation between the two
quantities.  Therefore we have repeated the same analysis with $Q$ fixed at its
theoretical prediction---as appears justified by the values for~$Q$ in
Table~\ref{tab:q_yt_fit}---in order to obtain more accurate estimates for
$y_{\rm t}^*$. The results, shown in the fifth column of
Table~\ref{tab:q_yt_fit}, are indeed in good agreement with the theoretically
expected values (last column). Thus, we have kept the thermal exponent fixed at
its theoretical value in the further analysis, just as in
Ref.~\onlinecite{prl}. The corresponding results for $Q$ and~$K_{\rm c}$ are
shown in Table~\ref{tab:qfit}.  As discussed in Ref.~\onlinecite{prl}, over the
full range of $\sigma$ and $d$ the Monte Carlo results for $Q$ show good
agreement with the renormalization prediction, thus confirming the universality
of this quantity above the upper critical dimension. In comparison with the
estimates presented in Table~I of Ref.~\onlinecite{prl}, two minor remarks
apply. First, for $Q(d=3,\sigma=0.4)$ one decimal place too much was quoted,
suggesting a too high accuracy. Secondly we note that the newest result for
$K_{\rm c}(d=3,\sigma=1.2)$ deviates two standard deviations from the earlier
estimate.

\begin{figure}
\begin{center}
\leavevmode
\epsfxsize 10cm
\epsfbox{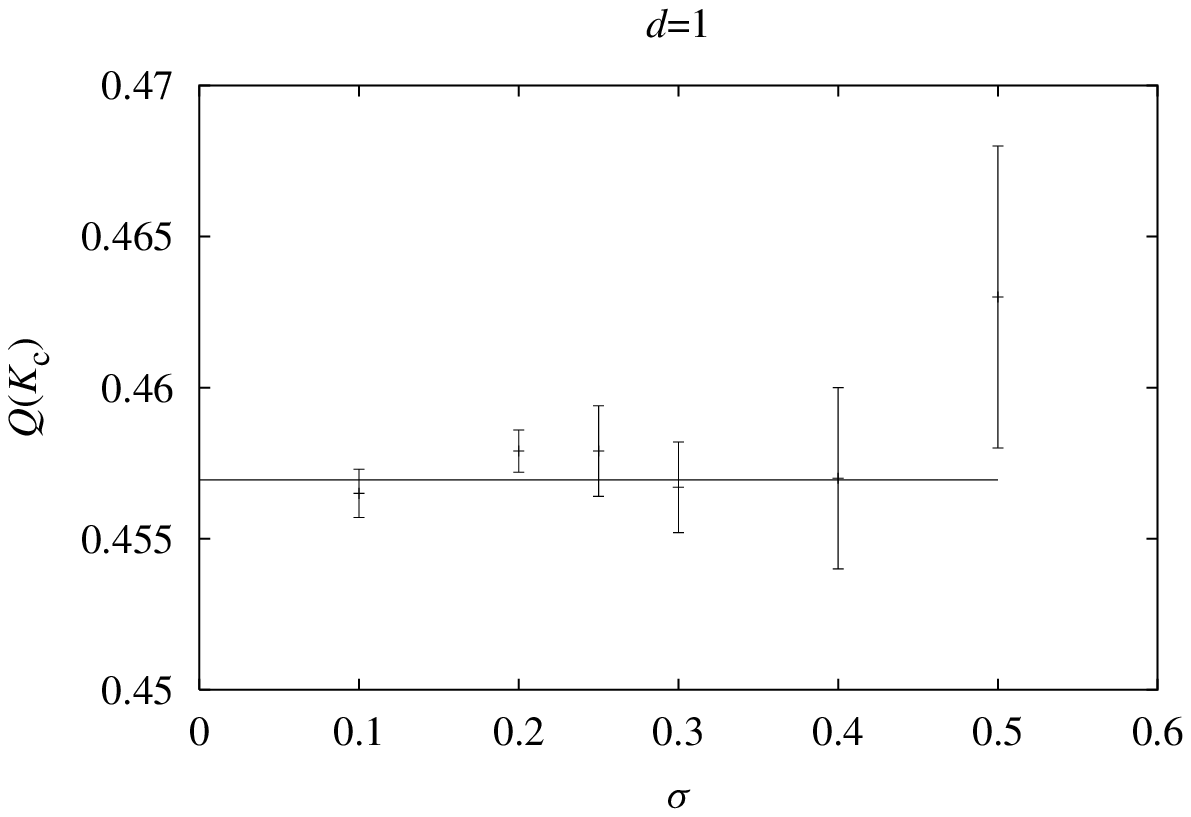} \\ 
\leavevmode
\epsfxsize 10cm
\epsfbox{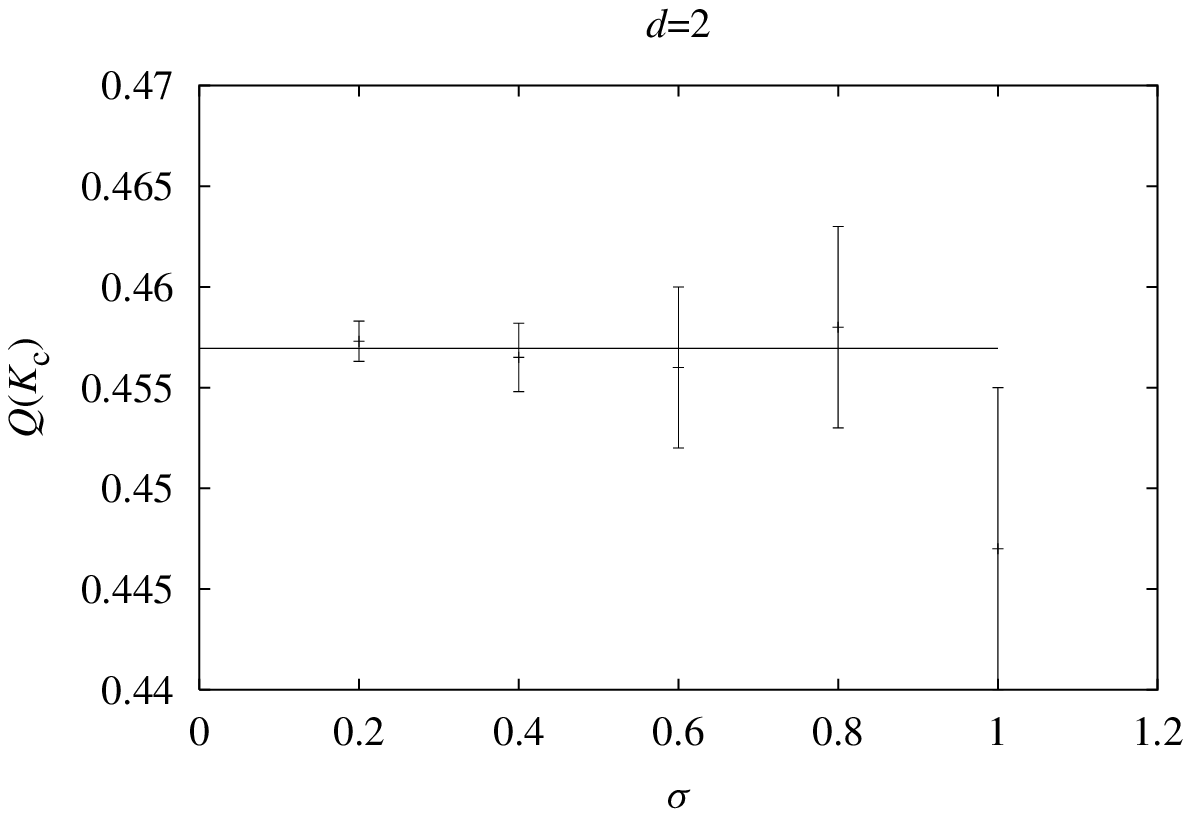} \\ 
\leavevmode
\epsfxsize 10cm
\epsfbox{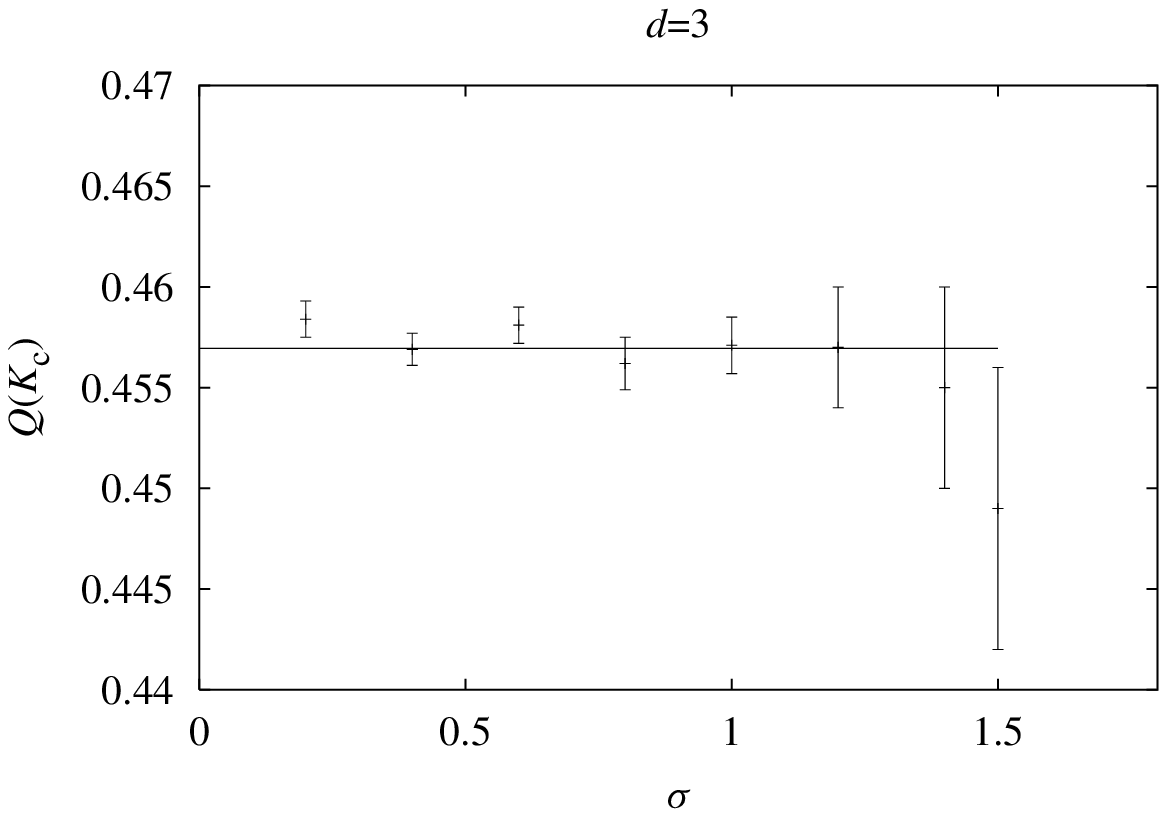}    
\end{center}
\caption{The amplitude ratio $Q$ as a function of the decay parameter $\sigma$
  in (a) $d=1$, (b) $d=2$, and (c) $d=3$ dimensions. The solid line marks the
  renormalization prediction.}
\label{fig:qplot}
\end{figure}

\begin{figure}
\begin{center}
\leavevmode
\epsfxsize 10cm
\epsfbox{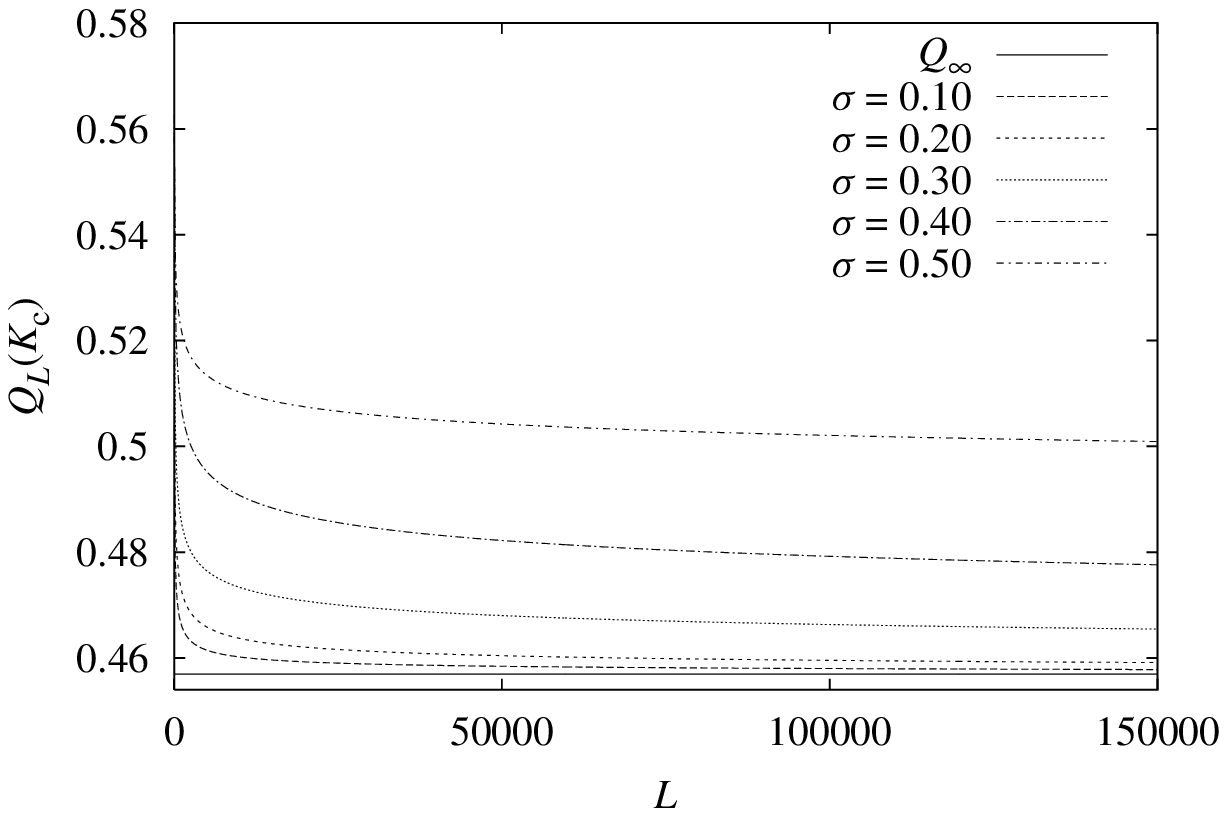} \\ 
\leavevmode
\epsfxsize 10cm
\epsfbox{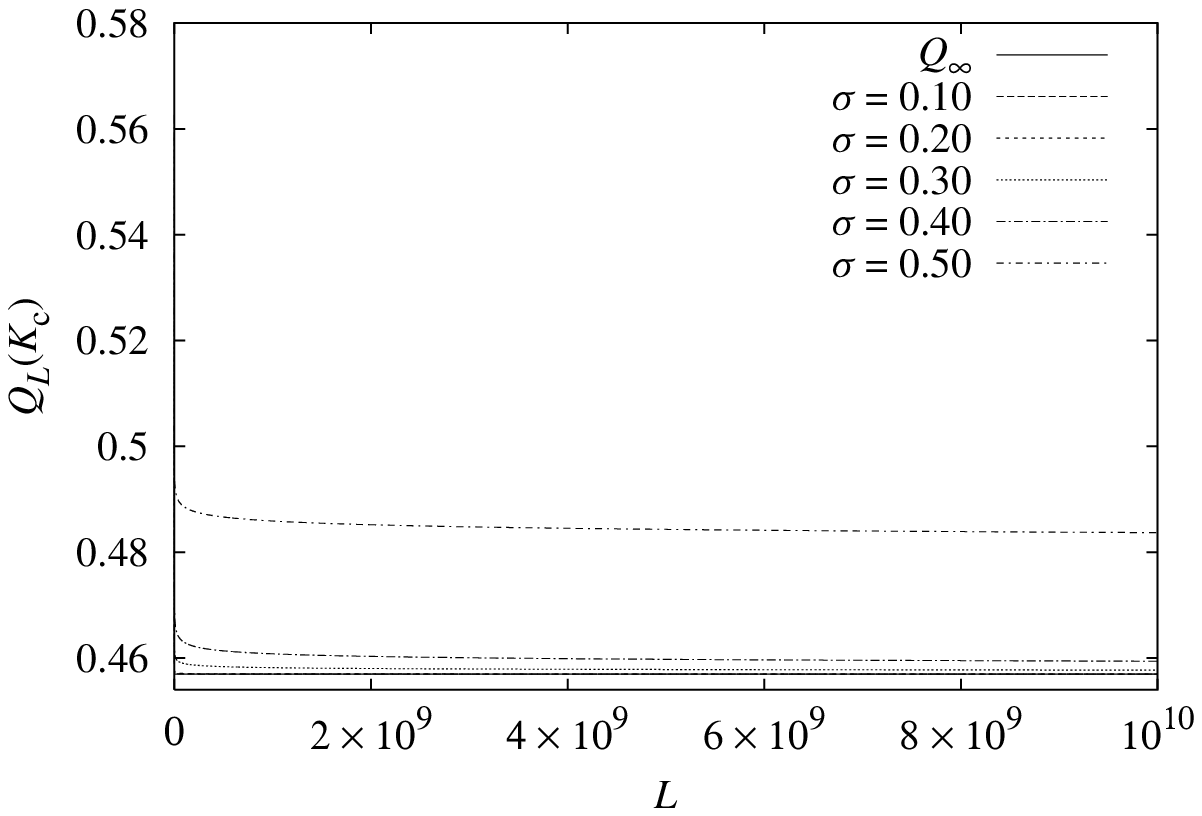}   
\end{center}
\caption{The amplitude ratio $Q$ in a one-dimensional system as a function of
  the system size $L$ for various values of $\sigma$. Figure~(a) illustrates
  the increase of the finite-size corrections when the upper critical dimension
  ($\sigma=d/2$) is approached. Figure~(b) emphasizes the difference between
  finite-size corrections {\em above\/} the upper critical dimension
  (power-law) and at the upper critical dimension itself (logarithmic).}
\label{fig:q-fss}
\end{figure}

The universality of~$Q$ is illustrated graphically in
Figs.~\ref{fig:qplot}(a)--\ref{fig:qplot}(c), where the increasing importance
of corrections to scaling upon approaching the upper critical dimension clearly
follows from the size of the error bars. At the upper critical dimension itself
($\epsilon=0$) this culminates in the appearance of logarithmic corrections,
where the finite-size scaling form of $Q_L$ is given by
\begin{eqnarray}
\label{eq:q-taylor-log}
 Q_L(T) &=& Q  + p_1 L^{y_{\rm t}}(\ln L)^{1/6} 
              \left[ t + v\frac{L^{-y_{\rm t}}}{(\ln L)^{2/3}} \right]
            + p_2 L^{2y_{\rm t}}(\ln L)^{1/3}
              \left[ t + v\frac{L^{-y_{\rm t}}}{(\ln L)^{2/3}} \right]^2
              \nonumber\\ 
        & & + q_1 L^{d-2y_{\rm h}} + \cdots + \frac{q_3}{\ln L} + \cdots \;.
\end{eqnarray}
The ellipses denote terms containing higher powers of $L^{d-2y_{\rm h}}$ and
$1/\ln L$. The extremely slow convergence of this series is reflected in the
uncertainty in the resulting estimates for~$Q$ at the upper critical dimension.
To illustrate the dependence of the finite-size corrections on $\epsilon$ more
directly, Fig.~\ref{fig:q-fss}(a) displays (for various values of $\sigma$) the
finite-size scaling functions as they follow from a least-squares fit of the
data for $d=1$ to Eqs.~(\ref{eq:q-taylor}) and~(\ref{eq:q-taylor-log}),
respectively. Although one clearly observes the increase of finite-size
corrections when $\sigma \rightarrow d/2$, the true nature of the logarithmic
corrections in~(\ref{eq:q-taylor-log}) cannot be appreciated from this graph.
To emphasize the difference between $\epsilon=0$ and $\epsilon<0$, we therefore
also show [Fig.~\ref{fig:q-fss}(b)] the same plot for the enormous range
$0<L<10^{10}$. Now it is evident how strongly the case $\epsilon=0$ differs
even from a case with strong power-law corrections, such as $\sigma=0.4$
($\epsilon=-0.2$).

We have used the universality of~$Q$ to considerably narrow the error margins
on $K_{\rm c}$ by fixing~$Q$ at its theoretical value in the least-squares fit.
The corresponding couplings are shown in Table~\ref{tab:qfit} as well. The
relative accuracy of the critical couplings lies between $1.5 \times 10^{-5}$
and $5.0 \times 10^{-5}$.  For the one-dimensional case, we can compare these
results to earlier estimates, see Table~\ref{tab:kc-comp}.  One notes that the
newest estimates are more than two orders of magnitude more accurate than
previous estimates. The first estimates~\cite{nagle70} were obtained by
carrying out exact calculations for chains of 1 to~20 spins and subsequently
extrapolating these results using Pad\'e approximants. Note that the estimates
for $T_{\rm c}$ in Ref.~\onlinecite{nagle70} are expressed in units of the
inverse of the Riemann zeta function and thus must be multiplied by
$\zeta(1+\sigma)$.  All couplings are somewhat too high, but still in fair
agreement with our estimates. The results of Doman~\cite{doman81} have no error
bars.  Still, his results are worrying, since he carries out a cluster
approach, obtaining critical couplings which start at the mean-field value for
cluster size zero and increase monotonically with increasing cluster size, as
they should, since mean-field theory yields a lower bound on the critical
couplings (see below).  Thus, he argues that the true couplings will lie {\em
  higher\/} than his best estimates (obtained for cluster size $10$).  However,
all these best estimates lie already {\em above\/} our estimates, which seems
to indicate a problem inherent in his approach.  Ref.~\onlinecite{glumac89}
presents results of an approximation coined ``finite-range scaling'' with error
margins of 1\%. For $\sigma=0.1$ the error is considerably underestimated, but
for the other values of the decay parameter the couplings agree with our
results well within the quoted errors.  The same technique was applied in
Ref.~\onlinecite{glumac93}, but now the uncertainty in the couplings was
estimated to be less than 10\%, for small $\sigma$ a few times larger. This is
clearly a too conservative estimate, as the difference with our results is only
a few percent for $\sigma=0.1$ and considerably less for larger $\sigma$. In
Ref.~\onlinecite{monroe90}, the coherent-anomaly method was used to obtain two
different estimates without error margins. We have quoted the average of the
two results, with their difference as a crude measure for the uncertainty. The
agreement is quite good, although all results lie systematically above our
values.  Yet another approach has been formulated in Ref.~\onlinecite{pires96},
where the Onsager reaction-field theory was applied to obtain a general
expression for the critical coupling,
\begin{equation}
 K_{\rm c}(\sigma) = \frac{\Gamma(1+\sigma) \sin (\pi \sigma /2)}%
                  {(1-\sigma)\pi^{1+\sigma}} \;.
\label{eq:orf}
\end{equation}
Unfortunately, no estimate for the accuracy of this expression is given, but it
seems to generally underestimate the critical coupling by a few percent.
Finally, some estimates have recently been obtained by means of the real-space
renormalization-group technique\@.~\cite{cannas96}

In addition, Monroe has calculated various bounds on the critical couplings as
shown in Table~\ref{tab:kc-bounds}.  The Bethe lattice
approximation~\cite{monroe92} was used to obtain both upper and lower bounds,
to which our results indeed conform, although it must be said that the upper
bounds do not constitute a very stringent criterion. Furthermore, the
application of Vigfusson's method~\cite{monroe94} has yielded even closer lower
bounds for $\sigma=0.1$ and $\sigma=0.2$.

\begin{figure}
\begin{center}
\leavevmode
\epsfxsize 10cm
\epsfbox{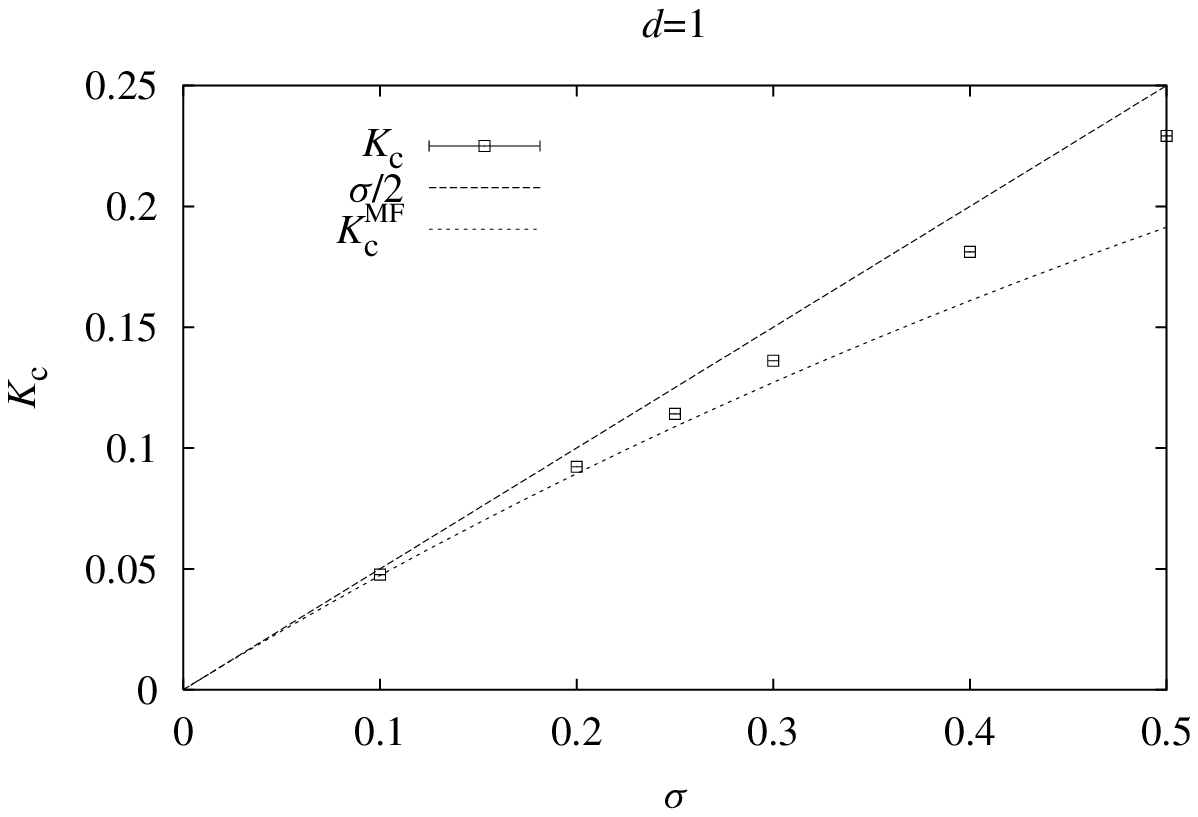} \\ 
\leavevmode
\epsfxsize 10cm
\epsfbox{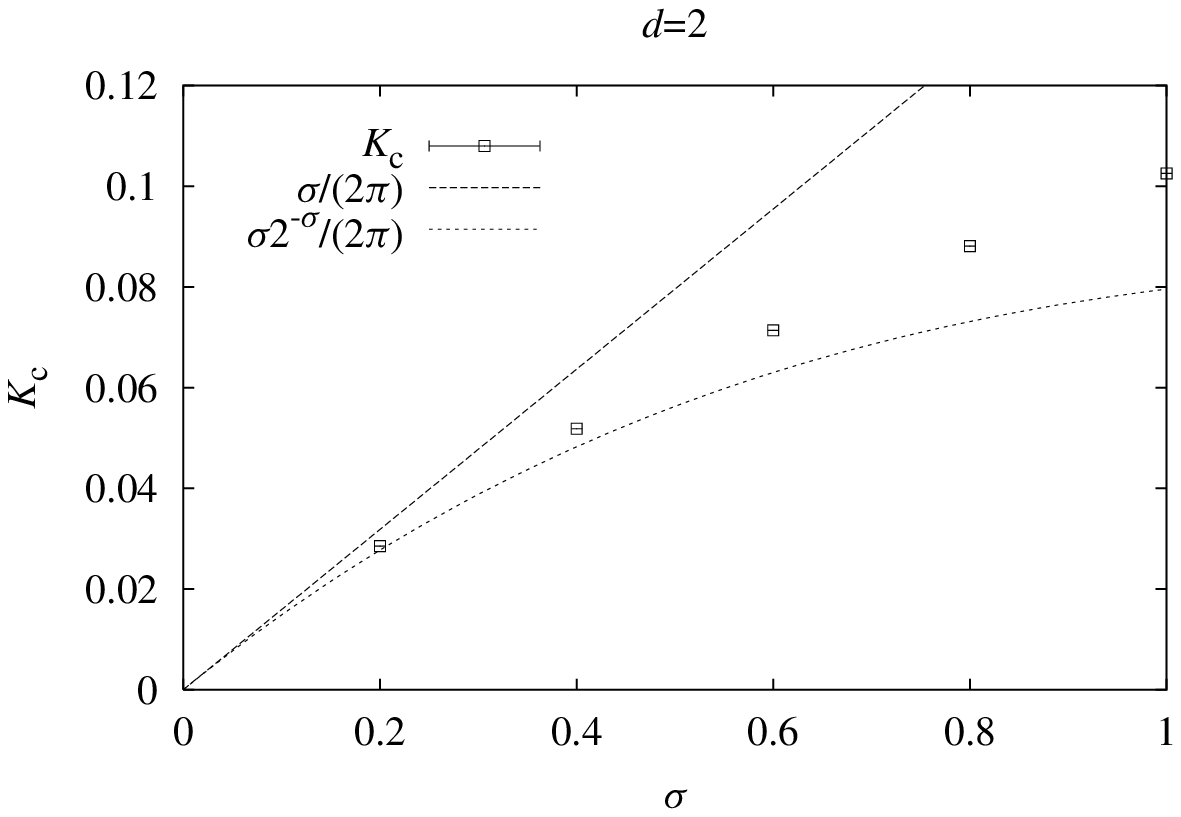} \\ 
\leavevmode
\epsfxsize 10cm
\epsfbox{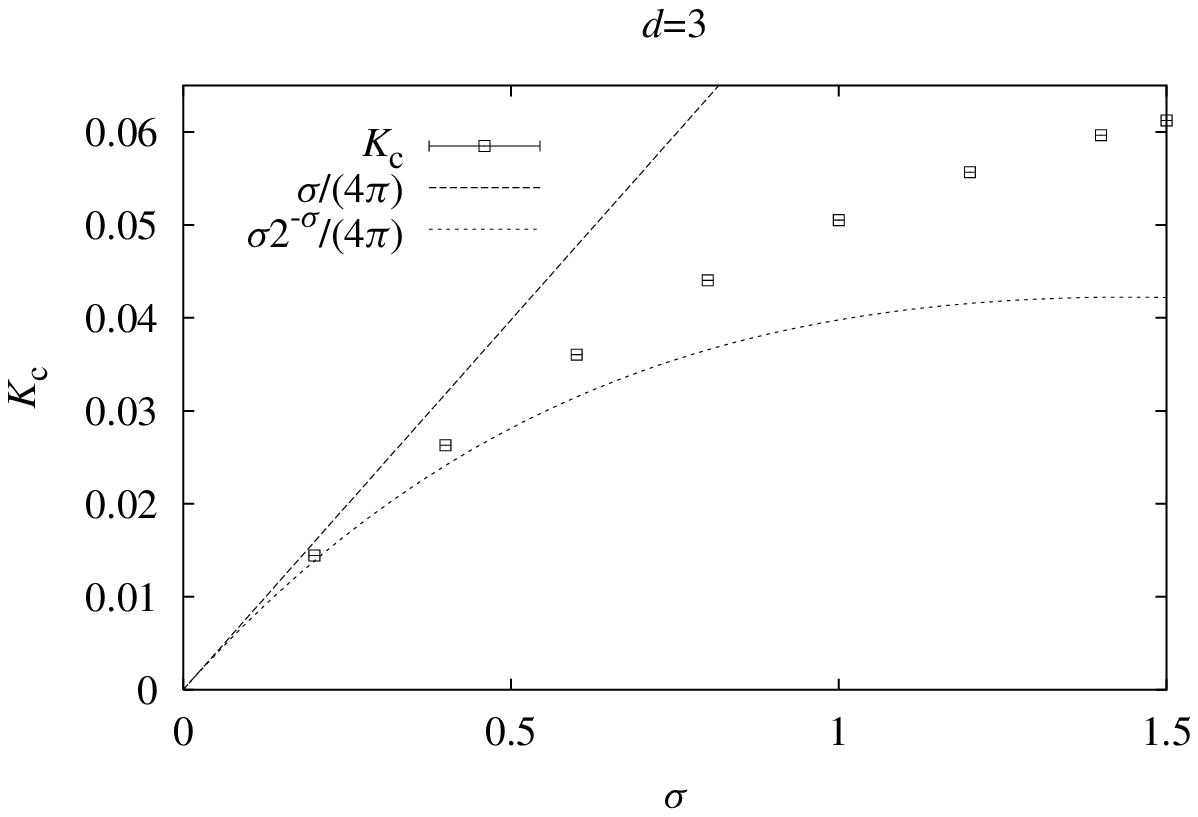}    
\end{center}
\caption{The critical coupling $K_{\rm c}$ as a function of the decay parameter
  $\sigma$ for (a) $d=1$, (b) $d=2$, and (c) $d=3$. Also shown is the
  asymptotic behavior for $\sigma \downarrow 0$ as predicted by mean-field
  theory and mean-field values for $K_{\rm c}$ over the full range of $0 <
  \sigma < d/2$ (for $d=2$ and $d=3$ only approximately).}
\label{fig:kc-sigma}
\end{figure}

Apart from these approximations, one may also use mean-field theory to make
some predictions concerning the critical coupling in the limit $\sigma
\downarrow 0$. It was shown by Brankov~\cite{brankov90} that in this limit the
$d$-dimensional system with an interaction potential $\propto \sigma /
r^{d+\sigma}$ is equivalent to the Husimi--Temperley mean spherical model.
More specifically, it was conjectured by Cannas~\cite{cannas95} that for the
one-dimensional case $\lim_{\sigma \to 0} K_{\rm c} \sim \sigma/2$, which is
also the first term in the Taylor expansion of Eq.~(\ref{eq:orf}).  Indeed, in
mean-field theory one has $zK_{\rm c}^{\rm MF}=1$, where $z$ is the
coordination number.  For $d=1$ this corresponds to the requirement
\begin{equation}
 2K_{\rm c}^{\rm MF}(\sigma) \sum_{n=1}^{\infty} \frac{1}{n^{1+\sigma}} 
    = 2K_{\rm c}^{\rm MF}(\sigma) \zeta(1+\sigma) = 1 \;,
\label{eq:riemann}
\end{equation}
where $\zeta(x)$ denotes the Riemann zeta function.  The expansion of
$\zeta(x)$ around $x=1$ yields the conjectured relation $\lim_{\sigma
  \downarrow 0} K_{\rm c}^{\rm MF} = \sigma/2$.  Figure~\ref{fig:kc-sigma}(a)
shows the critical coupling as a function of the decay parameter $\sigma$ along
with $K_{\rm c}^{\rm MF}(\sigma)$ and the asymptotic behavior for $\sigma
\downarrow 0$. One observes that $K_{\rm c}(\sigma)$ indeed approaches $K_{\rm
  c}^{\rm MF}(\sigma)$ when $\sigma$ approaches zero. Furthermore, $K_{\rm
  c}^{\rm MF}(\sigma)$ is smaller than $K_{\rm c}(\sigma)$ for all $\sigma$, as
one expects from the fact that mean-field theory {\em over\/}estimates the
critical temperature.  It is interesting to note that for $\sigma=0.1$ ($K_{\rm
  c}^{\rm MF} \approx 0.047239$) this lower bound already excludes the
estimates given in Refs.~\onlinecite{glumac93} and~\onlinecite{pires96} (cf.\ 
Table~\ref{tab:kc-comp}).  Replacing $zK_{\rm c}^{\rm MF}$ by the integrated
interaction, we can generalize such estimates to higher dimensionalities,
\begin{equation}
   K_{\rm c}^{\rm MF}(\sigma) \frac{2\pi^{d/2}}{\Gamma\left(\frac{d}{2}\right)}
   \int_{m_0}^\infty {\rm d}r\; \frac{1}{r^{1+\sigma}} = 1 \;.
\end{equation}
For $d>1$, the lower distance cutoff $m_0$ of the integral, i.e.\ the minimal
interaction distance with the nearest neighbors, does not have an isotropic
value, since there is no interaction within an elementary {\em cube\/} around
the origin.  Nevertheless, a constant value $m_0$, e.g.\ $m_0=1/2$, is a good
approximation.  Furthermore, for $d=1$ the integral is only a first-order
approximation of Eq.~(\ref{eq:riemann}), but for $d=2$ and $d=3$ it precisely
corresponds to the interaction~(\ref{eq:2d-j}) and its generalization to $d=3$,
respectively.  As a first estimate one thus obtains
\begin{equation}
\lim_{\sigma \downarrow 0} K_{\rm c}^{\rm MF}(\sigma) = 
   \frac{\Gamma\left(\frac{d}{2}\right)}{2\pi^{d/2}} \sigma m_0^\sigma \;.
\label{eq:kcmf}
\end{equation}
An expansion in terms of $\sigma$ shows that the first term is independent of
$m_0$. For $d=1,2,3$ one finds, respectively, $K_{\rm c}^{\rm MF} \sim
\sigma/2$, $K_{\rm c}^{\rm MF} \sim \sigma/(2\pi)$, $K_{\rm c}^{\rm MF} \sim
\sigma/(4\pi)$. Figures~\ref{fig:kc-sigma}(b) and~\ref{fig:kc-sigma}(c) show
$K_{\rm c}(\sigma)$ for $d=2$ and $d=3$, the corresponding asymptotes and
Eq.~(\ref{eq:kcmf}) with $m_0=1/2$.

The deviation of $K_{\rm c}(\sigma)$ from $K_{\rm c}^{\rm MF}(\sigma)$ is also
expressed by the last term in the renormalization expression~(\ref{eq:sol-r0}).
However, in order to assess the $\sigma$ dependence of this term one has to
calculate the $\sigma$ dependence of the coefficients $a$ and $c$, arising from
the integrals over the $\sigma$-dependent propagators.

\subsection{Determination of critical exponents}

\subsubsection{Magnetic susceptibility}
The magnetic susceptibility $\chi$ is directly proportional to the average
square magnetization density,
\begin{equation}
\chi = L^d \langle m^2 \rangle \;,
\end{equation}
and thus we can use Eq.~(\ref{eq:m-fss}) to analyze the finite-size data.
Expanding this equation in $t$ and $u$ we obtain for $\epsilon < 0$
\begin{equation}
\chi = L^{2y_{\rm h}^*-d} \left( a_0 +  a_1 \hat{t}L^{y_{\rm t}^*} 
     + a_2 \hat{t}^2 L^{2y_{\rm t}^*} + \cdots 
     + b_1 L^{y_{\rm i}} + \cdots \right)
\end{equation}
and for $\epsilon=0$
\begin{eqnarray}
\chi &=& L^{2y_{\rm h}-d} \sqrt{\ln L} \nonumber \\
 & & \times
   \left[ a_0 + a_1 L^{y_{\rm t}}(\ln L)^{1/6} 
          \left( t + v\frac{L^{-y_{\rm t}}}{(\ln L)^{2/3}} \right)
          + a_2 L^{2y_{\rm t}}(\ln L)^{1/3}
          \left( t + v\frac{L^{-y_{\rm t}}}{(\ln L)^{2/3}} \right)^2
          + \cdots + \frac{b_1}{\ln L} + \cdots 
   \right] \;.
\label{eq:chi-fss-du}
\end{eqnarray}
The analytic part of the free energy might give rise to an additional constant,
but this could not be observed in our simulations, because it is dominated by
the corrections to scaling.  In Table~\ref{tab:chi} we list the results of an
analysis of the numerical data. For all examined systems we have determined the
exponent $y_{\rm h}^*$ and the critical coupling. The estimates for the latter
are in good agreement with those obtained from the analysis of the universal
amplitude ratio~$Q$.  Furthermore, the exponents agree nicely, for all
dimensionalities, with the renormalization prediction $y_{\rm h}^*= 3d/4$. Just
as before, the uncertainties increase with increasing $\sigma$, although the
analyses at the upper critical dimension itself seem to yield better results
than those just above it. Compare in particular the results for $\sigma=1.4$
($y_{\rm i}=-0.2$) and $\sigma=1.5$. The logarithmic prefactor in
Eq.~(\ref{eq:chi-fss-du}) can be clearly observed in the sense that the quality
of the least-squares fit decreases considerably when this factor is omitted.
To reduce the uncertainty in the exponents we have repeated the analysis with
$K_{\rm c}$ fixed at the best values in Table~\ref{tab:qfit}, i.e.\ those
obtained with fixed~$Q$. The corresponding estimates of $y_{\rm h}^*$ are also
shown in Table~\ref{tab:chi} and are indeed in good agreement with the
renormalization predictions.

Now we can calculate the critical exponents and compare them to earlier
estimates for $d=1$. We do this for the correlation length exponent
$\nu=1/(y_{\rm t}^*+y_{\rm i}/2)$ and the magnetization exponent
$\beta=(d-y_{\rm h}^*)/y_{\rm t}^*$.  The results are shown in
Tables~\ref{tab:nu-comp} and~\ref{tab:beta-comp}.  Since all our estimates for
$y_{\rm t}^*$ and $y_{\rm h}^*$ agree with the renormalization values, also
$\nu$ and $\beta$ are in agreement with the classical critical exponents.
Unfortunately, the accuracy in both exponents is seriously hampered by the
uncertainty in $y_{\rm t}^*$, which has only been determined from the
temperature-dependent term in $Q$. In particular the results for $\nu$ from
Ref.~\onlinecite{glumac93} are, for small $\sigma$, in better agreement with
the theoretically predicted values than our estimates.  However, all previous
results, both for $\nu$ and for $\beta$, deviate seriously from the predicted
values when $\sigma$ approaches $1/2$, which is not the case for our values.
This can probably be attributed to the fact that corrections to scaling have
been taken into account more adequately.

\subsubsection{Spin--spin correlation function}
\begin{figure}
\begin{center}
\leavevmode
\epsfxsize 10cm
\epsfbox{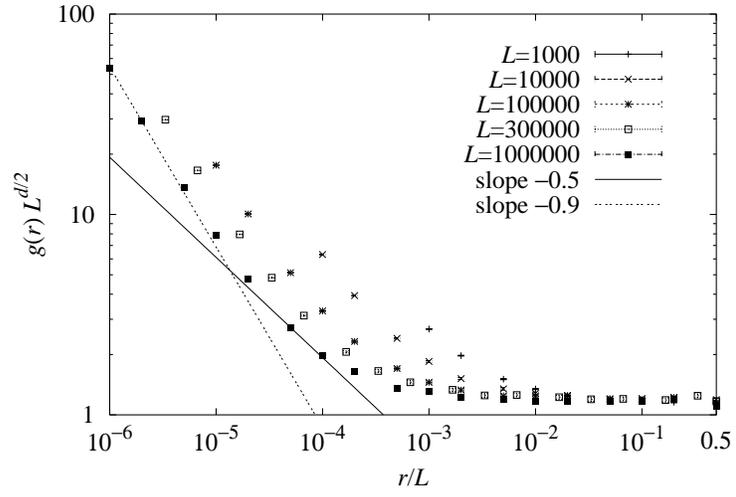} \\ 
\end{center}
\caption{The spin--spin correlation function versus $r/L$ in the
  one-dimensional model with $\sigma=0.1$. Results for various system sizes are
 shown. For a discussion see the text.}
\label{fig:g-decay}
\end{figure}

In Sec.~\ref{sec:fss} two different decay modes for the spin--spin correlation
function were derived. The relative magnitude of $r$ and $L$ determines which
of the modes applies.  In the bulk of our simulations we have restricted $r$ in
$g({\bf r})$ to $r=L/2$.  Since this quantity reflects the ${\bf k}={\bf 0}$
mode of the correlation function, we write for $\epsilon < 0$ an expression
analogous to that for the magnetic susceptibility,
\begin{equation}
 g(L/2) = L^{2y_{\rm h}^*-2d} \left[ c_0 +  c_1 \hat{t}L^{y_{\rm t}^*} 
          + c_2 \hat{t}^2 L^{2y_{\rm t}^*} + \cdots 
          + d_1 L^{y_{\rm i}} + \cdots \right]
\label{eq:g-fss}
\end{equation}
and for $\epsilon = 0$
\begin{eqnarray}
g(L/2) &=& L^{2y_{\rm h}-2d} \sqrt{\ln L} \nonumber\\
 & & \times  
   \left[ c_0 + c_1 L^{y_{\rm t}}(\ln L)^{1/6}
          \left( t + v\frac{L^{-y_{\rm t}}}{(\ln L)^{2/3}} \right)
          + c_2 L^{2y_{\rm t}}(\ln L)^{1/3}
          \left( t + v\frac{L^{-y_{\rm t}}}{(\ln L)^{2/3}} \right)^2
          + \cdots + \frac{d_1}{\ln L} + \cdots 
   \right] \;.
\label{eq:g-fss-du}
\end{eqnarray}
For values of~$r$ such that $g({\bf r})$ does {\em not\/} correspond to this
mode of the correlation function, the $\sigma$-dependent exponent $y_{\rm h}$
will appear in~(\ref{eq:g-fss}) instead of $y_{\rm h}^*$.  Furthermore, the
logarithmic prefactor in~(\ref{eq:g-fss-du}) will be absent, as it arises from
the dangerous irrelevant variable [cf.\ Eq.~(\ref{eq:m-fss-du})].  The results
of our analysis are shown in Table~\ref{tab:corrfunc}. They evidently
corroborate that the exponent $y_{\rm h}^*$ coincides with that appearing in
the susceptibility. Also the factor $\sqrt{\ln L}$ in~(\ref{eq:g-fss-du}) was
clearly visible in the least-squares analysis. The critical couplings agree
with the estimates from $Q$ and~$\chi$ and we have again tried to increase the
accuracy in $y_{\rm h}^*$ by repeating the analysis with $K_{\rm c}$ fixed at
their best values in Table~\ref{tab:qfit}. The accuracy of the results is
somewhat less than of those obtained from the magnetic susceptibility, because
we have now only used numerical data for even system sizes.  The fact that the
$L$ dependence of $g(L/2)$ is determined by the ${\bf k}={\bf 0}$ mode raises
the question whether one can also observe the power-law decay described by
$\eta$ in finite systems. To this end, we have sampled $g(r)$ as a function of
$r$ in the one-dimensional model. In order to clearly distinguish the two
predictions for the decay of $g(r)$ we have examined a system far from the
upper critical dimension, {\em viz.\/}\ with $\sigma=0.1$.  It turned out to be
necessary to sample {\em very\/} large system sizes to observe the regime where
$g(r) \propto r^{-(d-\sigma)}$.  Figure~\ref{fig:g-decay} displays the
spin--spin correlation function scaled with $L^{d/2}$ versus $r/L$. The scaling
makes the results collapse for $r$ of the order of the system size. Here, the
correlation function levels off. This is the mean-field like contribution to
the correlation function, which dominates in the spatial integral yielding the
magnetic susceptibility.  For small~$r$ the data do not collapse at all, which
shows that $g(r)$ exhibits different scaling behavior in this regime. Indeed,
the correlation function decays here as $r^{-(d-\sigma)}=r^{-0.9}$ and not as
$r^{-d/2}$. Note, however, that this regime is restricted to a small region of
$r$ and can only be observed for very large system sizes.

It is interesting to note that already Nagle and Bonner~\cite{nagle70} have
tried to calculate $\eta$ in a spin chain with long-range interactions from
finite-size data for the susceptibility. Because this calculation relied on the
assumption that $\chi(L,K_{\rm c})-\chi(L-1,K_{\rm c}) \sim g(L) \sim
L^{-(d-2+\eta)}$, they called the corresponding exponent $\tilde{\eta}$. The
results for $\tilde{\eta}$ turned out to assume a constant value approximately
equal to 1.50 for $0 < \sigma \leq 0.5$. Thus, the identification of
$\tilde{\eta}$ with $\eta$ was assumed to be invalid in
Ref.~\onlinecite{fisher72b}. Now we see that $\tilde{\eta}$ is in excellent
agreement with $\eta^* \equiv d+2-2y_{\rm h}^* = 2-d/2$.

\section{Conclusions}
\label{sec:conclusion}
In this paper we have studied systems with long-range interactions decaying as
$r^{-(d+\sigma)}$ in one, two, and three dimensions in the regime where these
interactions exhibit classical critical behavior, i.e., for $0 < \sigma \leq
d/2$. From the renormalization equations we have derived the scaling behavior,
including the corrections to scaling, for various quantities.  These
predictions, in particular the critical exponents and the scaling behavior of
the amplitude ratio $\langle m^2 \rangle ^2 / \langle m^4 \rangle$, have been
verified by accurate Monte Carlo results.  At the upper critical dimension, the
logarithmic factors appearing in the finite-size scaling functions could be
accurately observed.  The Monte Carlo results have been obtained with a
dedicated algorithm. This algorithm is many orders of magnitude faster (up to
the order of $10^8$ for the largest examined system) than a conventional Monte
Carlo algorithm for these systems. Our analysis has also yielded estimates for
the critical couplings. For $d=1$ these values have an accuracy which is more
than two orders of magnitude better than previous estimates and could thus
serve as a check for half a dozen different approximations. For $d=2$ and $d=3$
we have, to our best knowledge, obtained the first estimates for the critical
couplings.  Finally, we have given both theoretical and numerical arguments
that above the upper critical dimension the decay of the critical spin--spin
correlation function in finite systems consists of two regimes: One where it
decays as $r^{-(d-2+\eta)}$ and one where it is independent of the distance.

As an outlook we note that many interesting results may be expected {\em
below\/} the upper critical dimension, where neither any rigorous results nor
any accurate numerical results are available. This regime will be the subject
of a future investigation\@.\cite{nonclass}

\appendix
\section{Details of the Monte Carlo algorithm for long-range interactions}
The cluster algorithm applied in this study has been described for the first
time in Ref.~\onlinecite{ijmpc}. A somewhat more elaborate treatment of the
mathematical aspects was given in Ref.~\onlinecite{medran}. Although
conceptually no new aspects arise in the application to algebraically decaying
interactions in more than one dimension, several important practical issues
must be taken care of in actual simulations.  It is the purpose of this
Appendix to discuss these issues and their solutions in some more detail.  We
do not repeat the full cluster algorithm here, but only describe how the
cluster formation process proceeds from a given spin $s_i$ which has already
been added to the cluster (the so-called {\em current\/} spin).

As explained in Ref.~\onlinecite{ijmpc}, the key element of the algorithm lies
in splitting up the so-called bond-activation probability $p(s_i,s_j) =
\delta_{s_i s_j} p_{ij} = \delta_{s_i s_j} [1-\exp(-2J_{ij})]$ into two parts,
namely the Kronecker delta testing whether the spins $s_i$ and $s_j$ are
parallel and the ``provisional'' bond-activation probability $p_{ij}$. This
enables us to define a {\em cumulative bond probability} $C(k)$, from which we
can read off which bond is the next one to be provisionally activated,
\begin{equation}
 C(j) \equiv \sum_{n=1}^j P(n)
\label{eq:cumprob}
\end{equation}
with
\begin{equation}
 P(n) = (1-p_1)(1-p_2) \cdots (1-p_{n-1})p_n \;.
\end{equation}
$p_j \equiv 1-\exp(-2J_j)$ is an abbreviation for $p_{0j}$, i.e., we define the
origin at the position of the current spin. When comparing the expressions to
those in Ref.~\onlinecite{ijmpc} one must take into account that we now are
working with Ising instead of Potts couplings.  $P(n)$ is the probability that
in the first step $n-1$ bonds are skipped and the $n$th bond is provisionally
activated. Now the next bond $j$ that is provisionally activated is determined
by a random number $g \in [0,1 \rangle$: $j-1$ bonds are skipped if $C(j-1)
\leq g < C(j)$.  The number $j$ can be easily determined from $g$ once we have
tabulated the quantity $C(j)$ in a look-up table.  If the $j$th bond is placed
to a spin $s_j$ that is indeed parallel to the current spin $s_i$ then $s_j$ is
added to the cluster (i.e., the $j$th bond is activated).  Subsequently we skip
again a number of bonds before another bond at a distance $k>j$ is
provisionally activated.  The appropriate cumulative probability is now given
by a generalization of Eq.~(\ref{eq:cumprob}) (see Ref.~\onlinecite{ijmpc}),
\begin{equation}
 C_j(k) = \sum_{n=j+1}^{k} \left[ \prod_{m=j+1}^{n-1} (1-p_m) \right] p_n
        = 1 - \exp \left( -2 \sum_{n=j+1}^{k} J_n \right) \;.
\label{eq:gcumprob}
\end{equation}
In principle we need now for each value of $j$ another look-up table containing
the $C_j(k)$. This is hardly feasible and fortunately not necessary, as follows
from a comparison of Eqs.~(\ref{eq:cumprob}) and~(\ref{eq:gcumprob}). Namely,
\begin{equation}
 C(k) = C_0(k) = C(j) + \left[\prod_{i=1}^{j}(1-p_i) \right] C_j(k)
               = C(j) + \left[ 1 - C(j) \right] C_j(k)
\end{equation}
or $C_j(k) = [C(k)-C(j)]/[1-C(j)]$. So we can calculate $C_j(k)$ directly from
$C(k)$. In practice one realizes this by using the bond distance $j$ of the
previous bond that was provisionally activated to rescale the (new) random
number $g$ to $g' \in [C(j),1\rangle$; $g'=C(j)+[1-C(j)]g$.  Since we consider
only ferromagnetic interactions, $\lim_{j \to \infty} C(j)$ exists and is
smaller than 1, cf.\ Eq.~(\ref{eq:gcumprob}). Still we can accommodate only a
limited number of bond distances in our look-up table and must therefore devise
some approximation scheme to handle the tail of the long-range interaction,
which is essential for the critical behavior. This issue is addressed below.
Furthermore, this description only takes into account the bonds placed in one
direction. The actual implementation of the algorithm must of course allow for
bonds in both directions (assuming that $d=1$).

An alternative for the look-up table exists for interactions which can be
explicitly summed. In those cases, Eq.~(\ref{eq:gcumprob}) can be solved for
$k$, yielding an expression for the bond distance in terms of $C_j(k)$, i.e.,
in terms of the random number~$g$. For the interaction defined in
Sec.~\ref{sec:rigorous} the sum appearing in the right-hand side
of~(\ref{eq:gcumprob}) is (for $j=0$) the truncated Riemann zeta function,
\begin{equation}
 \sum_{n=1}^{k} J_n = K \sum_{n=1}^{k} \frac{1}{n^{d+\sigma}} \;,
\end{equation}
which cannot be expressed in closed form. In more than one dimension, a look-up
table is very impractical and an interaction which {\em can\/} be summed
explicitly becomes very desirable. Therefore we have taken an isotropic,
continuous interaction of the form $J=K/r^{d+\sigma}$. The interaction with a
spin at lattice site ${\bf n}$ is then given by the integral of $J$ over the
elementary square (cube) centered around ${\bf n}$ [cf.\ Eq.~(\ref{eq:2d-j})]
and the cumulative bond probability yields the (not necessarily integer-valued)
distance $k$ at which the first provisional bond is placed. To this end, the
sum in~(\ref{eq:gcumprob}) is replaced by a $d$-dimensional integral over the
coupling $J$. As $J$ is isotropic, only an integral over the radius remains,
which runs from the minimal bond distance up to $k$. Thus for $d=2$
Eq.~(\ref{eq:gcumprob}) reduces to
\begin{equation}
 C_j(k) = 1 - \exp \left[ -2 \frac{2\pi K}{\sigma} 
          \left( \frac{1}{j^\sigma} - \frac{1}{r^\sigma} \right) \right]
\label{eq:contcumprob}
\end{equation} 
and in $d=3$ the factor $2\pi$ is simply replaced by $4\pi$. Equating $C_j(k)$
to the random number $g$ we find
\begin{equation}
 k = \left[ j^{-\sigma} + \frac{\sigma}{4\pi K} \ln (1-g) \right] ^{-1/\sigma}
 \;.
\label{eq:bonddistance}
\end{equation}
Rescaling of the random number is no longer required: The lowest value, $g=0$,
leads to a provisional bond at the same distance as the previous one, $k=j$. If
$g=C_j(\infty) = 1-\exp[-(4\pi K/\sigma)j^{-\sigma}]$ the next provisional bond
lies at infinity and thus $g \in [C_j(\infty),1\rangle$ yields no bond at all.
Once the distance $k$ has been obtained, $d-1$ further random numbers $g_1,
g_2, \ldots$ are required to determine the {\em direction\/} of the bond. In
$d=2$, we set $\phi = g_1/(2\pi)$. The coordinates of the next provisional bond
(relative to the current spin) are then $(r_x,r_y) = (k\cos \phi,k\sin \phi)$,
which are rounded to the nearest integer coordinates. Finally, the periodic
boundary conditions are applied to map these coordinates onto a lattice site.
For the next provisional bond, $j$ is set equal to $k$ ({\em not\/} to the
rounded distance!)  and a new $k$ is determined. If no bond has been placed
yet, $j$ is set to $1/2$, the lowest possible bond distance. Hence it is
possible to find a $1/2 \leq k < \sqrt{2}/2$ and an angle $\phi$ such that the
corresponding lattice site is the origin. This does not affect the bond
probabilities, but it is of course a ``wasted'' Monte Carlo step. For $d=3$ the
process is similar, except that we need another random number $g_2$ to
determine a second angle $-\pi/2 < \psi \leq \pi/2$, such that $\sin\psi$ is
distributed uniformly; $\sin\psi = 1-2g_2$. The bond coordinates are given by
$(k\cos\psi \cos\phi, k\cos\psi \sin\phi, k\sin\psi)$.

This approach can also be applied in the one-dimensional case, where the
geometrical factor $2\pi$ in~(\ref{eq:contcumprob}) must be replaced by 2,
which reflects the fact that bonds can be put to the left and to the right of
the origin. The direction of the bond is then simply determined by another
random number.  As has already been mentioned in Ref.~\onlinecite{ijmpc}, this
can be used to cope with the limited size $M$ of the look-up table. Beyond the
bond distance $M$ the sum in~(\ref{eq:gcumprob}) is approximated by an
integral. I.e., if the random number $g$ lies in the interval
$[C(M),C(\infty)\rangle$, the bond distance $k$ is determined from the
one-dimensional version of~(\ref{eq:bonddistance}), where the lower part of the
integral is replaced by an explicit sum
\begin{equation}
 k = \left[ \left(M+\frac{1}{2}\right)^{-\sigma} + \sigma
  \left( \frac{1}{2K} \ln (1-g) + \sum_{n=1}^{M}\frac{1}{n^{1+\sigma}} \right) 
  \right]^{-1/\sigma} \;.
\label{eq:bonddist-1d}
\end{equation}
Here, the geometrical factor is absent, as we have opted to treat ``left'' and
``right'' separately in our simulations (no additional random number is
required in that case).  The approximation~(\ref{eq:bonddist-1d}) effectively
introduces a modification of the spin--spin interaction, which however can be
made arbitrarily small by increasing $M$.  Note that the offset $1/2$ in the
first term ensures a precise matching of the discrete sum and the integral
approximation: the random number $g=C(M)=1-\exp[-2K \sum_{n=1}^{M}
n^{-(1+\sigma)}]$ yields $k=M+1/2$ which is precisely the lowest $k$ that is
rounded to the integer bond distance $M+1$.

The accuracy of this procedure is further limited by the finite resolution of
random numbers. E.g., in our simulations the original random numbers are
integers in the range $[0,2^{32}-1]$. Thus, for bond distances $l$ such that
$C(l)-C(l-1)$ is of the order $2^{-32}$, the discreteness of the random numbers
is no longer negligible. For $d=2$ and $d=3$, the discreteness of the angles
also limits the lattice sites that can be selected for a provisional bond, but
this generally occurs at distances larger than $l$.  Once the value of $l$ has
been determined, with a safe margin, there are various approaches to this
limitation. One may, e.g., draw another random number to determine the precise
bond distance. A simpler approach is to distribute all bonds beyond $l$
uniformly over the lattice, in order to prevent that certain lattice sites are
never selected. However, one should take care that such simple approaches do
not essentially modify the critical behavior. If $l$ is relatively small, the
error introduced by a random distribution of the bond distances might be larger
than the effect of an interaction which decreases slightly nonmonotonically at
large distances. Furthermore, in order to preserve the symmetry
of the lattice, such a uniform distribution of the bonds should occur outside a
square (cube) instead of a circle (sphere) with radius~$l$.

\newpage

\newpage

\newpage

\begin{table}
\caption{The amplitude ratio $Q$ and the thermal exponent $y_{\rm t}^*$ for
  systems with long-range interactions in one, two, and three dimensions, for
  several values of the decay parameter $0 < \sigma \leq d/2$. The values in
  the fifth column have been obtained with $Q$ fixed at the theoretically
  predicted value (see text) and the last column lists the renormalization
  predictions for $y_{\rm t}^*$.}
\label{tab:q_yt_fit}
\renewcommand{\arraystretch}{1.15}
\begin{tabular}{c|d|d|d|d|c}
$d$ & $\sigma$ & $Q$         & $y_{\rm t}^*$ & $y_{\rm t}^*$ & RG \\
\hline
1   & 0.1      & 0.4566 (8)  & 0.507 (7)     & 0.507 (7)     & $\frac{1}{2}$ \\
1   & 0.2      & 0.455 (4)   & 0.54 (4)      & 0.504 (12)    & $\frac{1}{2}$ \\
1   & 0.25     & 0.457 (3)   & 0.500 (8)     & 0.500 (5)     & $\frac{1}{2}$ \\
1   & 0.3      & 0.454 (2)   & 0.519 (14)    & 0.506 (12)    & $\frac{1}{2}$ \\
1   & 0.4      & 0.457 (3)   & 0.50 (2)      & 0.50 (2)      & $\frac{1}{2}$ \\
1   & 0.5      & 0.462 (6)   & 0.51 (5)      & 0.49 (2)      & $\frac{1}{2}$ \\
\hline
2   & 0.2      & 0.4574 (10) & 1.01 (2)      & 1.01 (2)      & 1 \\
2   & 0.4      & 0.455 (2)   & 1.02 (2)      & 1.009 (15)    & 1 \\
2   & 0.6      & 0.450 (6)   & 1.04 (4)      & 1.008 (17)    & 1 \\
2   & 0.8      & 0.454 (6)   & 1.03 (9)      & 1.03 (3)      & 1 \\
2   & 1.0      & 0.450 (10)  & 1.02 (3)      & 1.03 (2)      & 1 \\
\hline
3   & 0.2      & 0.4581 (11) & 1.51 (3)      & 1.513 (18)    & $\frac{3}{2}$ \\
3   & 0.4      & 0.4561 (10) & 1.521 (18)    & 1.512 (15)    & $\frac{3}{2}$ \\
3   & 0.6      & 0.453 (3)   & 1.53 (4)      & 1.521 (14)    & $\frac{3}{2}$ \\
3   & 0.8      & 0.458 (2)   & 1.48 (2)      & 1.487 (10)    & $\frac{3}{2}$ \\
3   & 1.0      & 0.453 (10)  & 1.52 (7)      & 1.508 (9)     & $\frac{3}{2}$ \\
3   & 1.2      & 0.447 (8)   & 1.56 (2)      & 1.519 (10)    & $\frac{3}{2}$ \\
3   & 1.4      & 0.454 (5)   & 1.48 (3)      & 1.48 (3)      & $\frac{3}{2}$ \\
3   & 1.5      & 0.449 (8)   & 1.53 (5)      & 1.46 (3)      & $\frac{3}{2}$
\end{tabular}
\end{table}

\begin{table}
\caption{The amplitude ratio $Q$ and critical couplings $K_{\rm c}$ for systems
  with long-range interactions in one, two, and three dimensions, for several
  values of the decay parameter $0 < \sigma \leq d/2$. The thermal exponent
  (see Table~\protect\ref{tab:q_yt_fit}) was kept fixed at its theoretical
  value in all analyses.  The estimates for $K_{\rm c}$ in the last column have
  been obtained by fixing $Q$ at its renormalization prediction. The numbers
  between parentheses represent the errors in the last decimal places.}
\label{tab:qfit}
\begin{tabular}{c|d|d|d|d}
$d$ & $\sigma$ & $Q$         & $K_{\rm c}$     & $K_{\rm c}$    \\
\hline
1   & 0.1      & 0.4565 (8)  & 0.0476162 (13)  & 0.0476168 (6)  \\ 
1   & 0.2      & 0.4579 (7)  & 0.092234 (2)    & 0.0922314 (15) \\ 
1   & 0.25     & 0.4579 (15) & 0.114143 (4)    & 0.1141417 (19) \\ 
1   & 0.3      & 0.4567 (15) & 0.136113 (4)    & 0.136110 (2)   \\ 
1   & 0.4      & 0.457 (3)   & 0.181151 (8)    & 0.181150 (3)   \\ 
1   & 0.5      & 0.463 (5)   & 0.229157 (8)    & 0.229155 (6)   \\ 
\hline
2   & 0.2      & 0.4573 (10) & 0.028533 (3)    & 0.0285324 (14) \\ 
2   & 0.4      & 0.4565 (17) & 0.051824 (4)    & 0.0518249 (14) \\ 
2   & 0.6      & 0.456 (4)   & 0.071364 (7)    & 0.071366  (2)  \\ 
2   & 0.8      & 0.458 (5)   & 0.088094 (7)    & 0.088094  (2)  \\ 
2   & 1.0      & 0.447 (8)   & 0.102556 (5)    & 0.102558  (5)  \\ 
\hline
3   & 0.2      & 0.4584 (9)  & 0.0144361 (10)  & 0.0144354 (6)  \\ 
3   & 0.4      & 0.4569 (8)  & 0.0262927 (16)  & 0.0262929 (7)  \\ 
3   & 0.6      & 0.4581 (9)  & 0.036050 (2)    & 0.0360469 (11) \\ 
3   & 0.8      & 0.4562 (13) & 0.044034 (2)    & 0.0440354 (10) \\ 
3   & 1.0      & 0.4571 (14) & 0.050515 (2)    & 0.0505152 (12) \\ 
3   & 1.2      & 0.457 (3)   & 0.055682 (3)    & 0.0556825 (14) \\ 
3   & 1.4      & 0.455 (5)   & 0.059666 (2)    & 0.0596669 (11) \\ 
3   & 1.5      & 0.449 (7)   & 0.061251 (2)    & 0.061253 (2)      
\end{tabular}
\end{table}

\begin{table}
\caption{Comparison between our best estimates of the critical couplings
 $K_{\rm c}$ for the one-dimensional system and earlier estimates.}
\label{tab:kc-comp}
\begin{tabular}{l|l|l|l|l|l|l|l|l}
$\sigma$ & This work & Ref.~\protect\onlinecite{nagle70} & 
 Ref.~\protect\onlinecite{doman81}  & Ref.~\protect\onlinecite{glumac89} &
 Ref.~\protect\onlinecite{glumac93} & Ref.~\protect\onlinecite{monroe90} &
 Ref.~\protect\onlinecite{pires96}  &
 Ref.~\protect\onlinecite{cannas96} \\
\hline
0.1      & 0.0476168 (6)  & ---            & 0.0478468 & 0.0505 (5)  & 0.04635
         & 0.04777 (12)   & 0.0469         & 0.0481 \\ 
0.2      & 0.0922314 (15) & 0.0926 (5)     & 0.0933992 & 0.0923 (9)  & 0.09155
         & 0.0928 (3)     & 0.0898         & ---    \\
0.25     & 0.1141417 (19) & ---            & ---       & ---         & ---
         & ---            & 0.1106         & ---    \\
0.3      & 0.136110 (2)   & 0.1370 (7)     & 0.138478  & 0.1362 (14) & 0.1359
         & 0.1375 (10)    & 0.1314         & 0.144  \\    
0.4      & 0.181150 (3)   & 0.1825 (10)    & 0.184081  & 0.1815 (18) & 0.1813
         & 0.183 (2)      & 0.1750         & ---    \\
0.5      & 0.229155 (6)   & 0.2307 (14)    & 0.230821  & 0.230 (2)   & 0.2295
         & 0.231 (4)      & 0.2251         & 0.250
\end{tabular}
\end{table}

\begin{table}
\caption{Comparison of our best estimates of the critical couplings for the
 one-dimensional system with some lower and upper bounds.}
\label{tab:kc-bounds}
\begin{tabular}{l|l|l|l|l}
$\sigma$ & This work & 
 Ref.~\protect\onlinecite{monroe92} &
 Ref.~\protect\onlinecite{monroe92} &
 Ref.~\protect\onlinecite{monroe94} \\ \hline
0.1 & 0.0476168 (6)  & $\geq$ 0.04726 & $\leq$ 0.09456 & $\geq$ 0.04753 \\
0.2 & 0.0922314 (15) & $\geq$ 0.08947 & $\leq$ 0.1792  & $\geq$ 0.09162 \\
0.3 & 0.136110 (2)   & $\geq$ 0.1273  & $\leq$ 0.2558  & ---            \\
0.4 & 0.181150 (3)   & $\geq$ 0.1615  & $\leq$ 0.3258  & ---            \\
0.5 & 0.229155 (6)   & $\geq$ 0.1923  & $\leq$ 0.3903  & ---
\end{tabular}
\end{table}

\begin{table}
\caption{Estimates for the critical coupling $K_{\rm c}$ and the exponent
  $y_{\rm h}^*$ as obtained from the analysis of the magnetic susceptibility.
  The values for $y_{\rm h}^*$ in the fifth column have been obtained by fixing
  $K_{\rm c}$ at their best estimates from Table~\protect\ref{tab:qfit}; the
  error margins do not include the uncertainty in these values for $K_{\rm
  c}$.}
\label{tab:chi}
\renewcommand{\arraystretch}{1.15}
\begin{tabular}{c|d|d|d|d|c}
$d$ & $\sigma$ & $K_{\rm c}$    & $y_{\rm h}^*$ & $y_{\rm h}^*$ & RG        \\
\hline
1   & 0.1      & 0.0476161 (19) & 0.7487 (14) & 0.7493 (6)  & $\frac{3}{4}$ \\
1   & 0.2      & 0.092239 (4)   & 0.752 (2)   & 0.7504 (10) & $\frac{3}{4}$ \\
1   & 0.25     & 0.114145 (4)   & 0.7477 (15) & 0.7468 (16) & $\frac{3}{4}$ \\
1   & 0.3      & 0.136110 (5)   & 0.747 (3)   & 0.7490 (17) & $\frac{3}{4}$ \\
1   & 0.4      & 0.181170 (10)  & 0.749 (5)   & 0.746 (3)   & $\frac{3}{4}$ \\
1   & 0.5      & 0.229153 (6)   & 0.748 (2)   & 0.7490 (8)  & $\frac{3}{4}$ \\
\hline
2   & 0.2      & 0.028537 (5)   & 1.500 (6)   & 1.495 (3)   & $\frac{3}{2}$ \\
2   & 0.4      & 0.051830 (6)   & 1.498 (9)   & 1.496 (3)   & $\frac{3}{2}$ \\
2   & 0.6      & 0.071370 (5)   & 1.497 (6)   & 1.498 (2)   & $\frac{3}{2}$ \\
2   & 0.8      & 0.088095 (10)  & 1.496 (5)   & 1.495 (3)   & $\frac{3}{2}$ \\
2   & 1.0      & 0.102556 (3)   & 1.495 (4)   & 1.497 (2)   & $\frac{3}{2}$ \\
\hline
3   & 0.2      & 0.0144347 (9)  & 2.249 (2)   & 2.2504 (8)  & $\frac{9}{4}$ \\
3   & 0.4      & 0.026296 (2)   & 2.250 (6)   & 2.246 (3)   & $\frac{9}{4}$ \\
3   & 0.6      & 0.036046 (3)   & 2.246 (7)   & 2.244 (5)   & $\frac{9}{4}$ \\
3   & 0.8      & 0.0440349 (17) & 2.243 (4)   & 2.246 (3)   & $\frac{9}{4}$ \\
3   & 1.0      & 0.050516 (3)   & 2.239 (2)   & 2.243 (7)   & $\frac{9}{4}$ \\
3   & 1.2      & 0.055679 (2)   & 2.247 (11)  & 2.251 (7)   & $\frac{9}{4}$ \\
3   & 1.4      & 0.0596636 (18) & 2.27 (3)    & 2.26 (2)    & $\frac{9}{4}$ \\
3   & 1.5      & 0.061251 (2)   & 2.257 (12)  & 2.249 (5)   & $\frac{9}{4}$
\end{tabular}
\end{table}

\begin{table}
\caption{The correlation length exponent $\nu$ as a function of $\sigma$ for
  the one-dimensional model, together with earlier estimates and the
  renormalization predictions.}
\label{tab:nu-comp}
\begin{tabular}{d|d|d|d|d|d}
$\sigma$ & This work  
   & Ref.~\protect\onlinecite{glumac89}
   & Ref.~\protect\onlinecite{glumac93}
   & Ref.~\protect\onlinecite{cannas96} 
   & RG \\ \hline
0.1      & 9.3 (6)   & 9.12 & 9.9  & 10.48  & 10.0  \\
0.2      & 4.9 (3)   & 4.90 & 4.95  & ---    & 5.0   \\     
0.25     & 4.00 (8)  & ---  & ---   & ---    & 4.0   \\ 
0.3      & 3.27 (12) & 3.41 & 3.32  & 3.90   & 3.3$\ldots$ \\
0.4      & 2.50 (13) & 2.71 & 2.68  & ---    & 2.5   \\   
0.5      & 2.04 (8)  & 2.34 & 2.33  & 2.81   & 2.0   
\end{tabular}
\end{table}

\begin{table}
\caption{The magnetization exponent $\beta$ as a function of $\sigma$ for the
   one-dimensional model, together with earlier estimates and the
   renormalization predictions.}
\label{tab:beta-comp}
\renewcommand{\arraystretch}{1.15}
\begin{tabular}{d|d|d|d|d}
$\sigma$ & This work  
   & Ref.~\protect\onlinecite{nagle70}
   & Ref.~\protect\onlinecite{monroe90}
   & RG \\ \hline
0.1  & 0.494 (8)  & ---  & 0.495 & $\frac{1}{2}$ \\
0.2  & 0.495 (13) & 0.5  & 0.482 & $\frac{1}{2}$ \\
0.25 & 0.506 (8)  & ---  & ---   & $\frac{1}{2}$ \\
0.3  & 0.497 (15) & 0.48 & 0.460 & $\frac{1}{2}$ \\
0.4  & 0.51 (2)   & 0.45 & 0.435 & $\frac{1}{2}$ \\
0.5  & 0.51 (2)   & 0.39 & 0.408 & $\frac{1}{2}$
\end{tabular}
\end{table}

\begin{table}
\caption{Estimates for the critical coupling $K_{\rm c}$ and the exponent
  $y_{\rm h}^*$ as obtained from the analysis of the spin--spin correlation
  function.  The values for $y_{\rm h}^*$ in the fifth column have been
  obtained by fixing $K_{\rm c}$ at their best estimates from
  Table~\protect\ref{tab:qfit}; the error margins do not include the
  uncertainty in these values for $K_{\rm c}$.}
\label{tab:corrfunc}
\renewcommand{\arraystretch}{1.15}
\begin{tabular}{c|d|d|d|d|c}
$d$ & $\sigma$ & $K_{\rm c}$    & $y_{\rm h}^*$ & $y_{\rm h}^*$ & RG        \\
\hline
1   & 0.1      & 0.047619 (3)   & 0.750 (2)   & 0.7488 (9)  & $\frac{3}{4}$ \\
1   & 0.2      & 0.092233 (7)   & 0.749 (3)   & 0.7513 (16) & $\frac{3}{4}$ \\
1   & 0.25     & 0.114148 (10)  & 0.750 (5)   & 0.747 (2)   & $\frac{3}{4}$ \\
1   & 0.3      & 0.136116 (7)   & 0.753 (5)   & 0.752 (3)   & $\frac{3}{4}$ \\
1   & 0.4      & 0.181158 (15)  & 0.747 (7)   & 0.750 (4)   & $\frac{3}{4}$ \\
1   & 0.5      & 0.229150 (7)   & 0.749 (2)   & 0.7503 (10) & $\frac{3}{4}$ \\
\hline
2   & 0.2      & 0.028535 (7)   & 1.499 (9)   & 1.496 (3)   & $\frac{3}{2}$ \\
2   & 0.4      & 0.051831 (6)   & 1.505 (6)   & 1.499 (4)   & $\frac{3}{2}$ \\
2   & 0.6      & 0.071369 (6)   & 1.507 (4)   & 1.502 (4)   & $\frac{3}{2}$ \\
2   & 0.8      & 0.088091 (6)   & 1.495 (7)   & 1.497 (3)   & $\frac{3}{2}$ \\
2   & 1.0      & 0.102554 (4)   & 1.490 (6)   & 1.496 (3)   & $\frac{3}{2}$ \\
\hline
3   & 0.2      & 0.0144348 (16) & 2.256 (6)   & 2.254 (4)   & $\frac{9}{4}$ \\
3   & 0.4      & 0.026296 (3)   & 2.257 (8)   & 2.245 (5)   & $\frac{9}{4}$ \\
3   & 0.6      & 0.036053 (4)   & 2.262 (10)  & 2.246 (4)   & $\frac{9}{4}$ \\
3   & 0.8      & 0.044035 (4)   & 2.252 (11)  & 2.250 (5)   & $\frac{9}{4}$ \\
3   & 1.0      & 0.050511 (5)   & 2.228 (15)  & 2.249 (9)   & $\frac{9}{4}$ \\
3   & 1.2      & 0.055680 (3)   & 2.253 (14)  & 2.257 (9)   & $\frac{9}{4}$ \\
3   & 1.4      & 0.059667 (2)   & 2.22 (4)    & 2.31 (4)    & $\frac{9}{4}$ \\
3   & 1.5      & 0.061251 (5)   & 2.26 (3)    & 2.248 (7)   & $\frac{9}{4}$
\end{tabular}
\end{table}

\end{document}